\documentclass[12pt]{iopart}
\usepackage[utf8]{inputenc}

\usepackage{graphicx}
\makeatletter
\expandafter\let\csname equation*\endcsname\relax
\expandafter\let\csname endequation*\endcsname\relax
\makeatother
\usepackage{amsmath}

\usepackage{float}

\usepackage[section]{placeins} 

\usepackage[table]{xcolor}

\usepackage[euler]{textgreek}
\usepackage{bm}

\usepackage[breaklinks=true,colorlinks=true,linkcolor=blue,urlcolor=blue,citecolor=blue,hyperfootnotes=false]{hyperref}

\usepackage{amssymb}

\usepackage{subcaption}
\makeatletter
\renewcommand\p@subfigure{\thefigure-}
\makeatother

\usepackage[numbers,sort&compress]{natbib}
\begin{document}

\title{Characterization of ELM Pacing via Vertical Jogs on DIII-D}
\makeatletter
\let\orig@fnsymbol\@fnsymbol
\renewcommand\@fnsymbol[1]{*}
\makeatother
\author{K Yasoda$^{1,}$\footnote{Authors to whom any correspondence should be addressed.}, D Panici$^1$, A O Nelson$^2$, F M Laggner$^3$, S K Kim$^4$ and E Kolemen$^{1,4,}$\footnotemark[1]}
\address{$^1$ Princeton University, Princeton, NJ 08544, USA}
\address{$^2$ Columbia University, New York, NY 10027, USA}
\address{$^3$ North Carolina State University, Raleigh, NC 27695, USA}
\address{$^4$ Princeton Plasma Physics Laboratory, Princeton, NJ 08543, USA}
\ead{keiyasoda@princeton.edu; ekolemen@pppl.gov, ekolemen@princeton.edu}

\begin{abstract}
Edge localized mode (ELM) pacing via vertical plasma oscillations or jogging has been successfully demonstrated on DIII-D. Rapid vertical movement of the plasma toward the X-point has been shown to effectively trigger ELMs. By vertically oscillating the plasma at a rate of 10~Hz, the ELM frequency increased from $\sim$5~Hz, the natural ELM frequency in similar DIII-D discharges, to 20~Hz. Downward jogs have been observed to trigger multiple ELMs in one cycle. ELMs triggered at higher than natural frequencies lead to smaller decreases in stored energy, from ~8\% to as little as below 1\%. As a consequence, the peak heat flux to the divertor has been observed to be reduced by a factor of $\sim$2. In addition, a reduction in the carbon impurity concentration has been observed. During downward jogs in the lower single null (LSN) configuration, the X-point movement is slower and smaller than the top of the plasma. As a result, a reduction in the plasma cross-section and hence volume has been observed. To understand the mechanism of ELM triggering by jogging, a toy model of the edge toroidal current has been built and tested with DIII-D experiment data. The experimental data and model suggest that when the plasma moves down toward the X-point, a net positive toroidal current is locally induced in the edge region. ELITE stability analysis suggests that this current pushes the plasma state across the peeling side of the peeling-ballooning stability boundary into the unstable region triggering ELMs.

\end{abstract}

\noindent{\it Keywords}: ELM pacing, vertical oscillations, ELMs, pedestal

\maketitle
\makeatletter\let\@fnsymbol\orig@fnsymbol\makeatother

\begin{figure}[htbp]
    \centering
    \includegraphics[width=\textwidth]{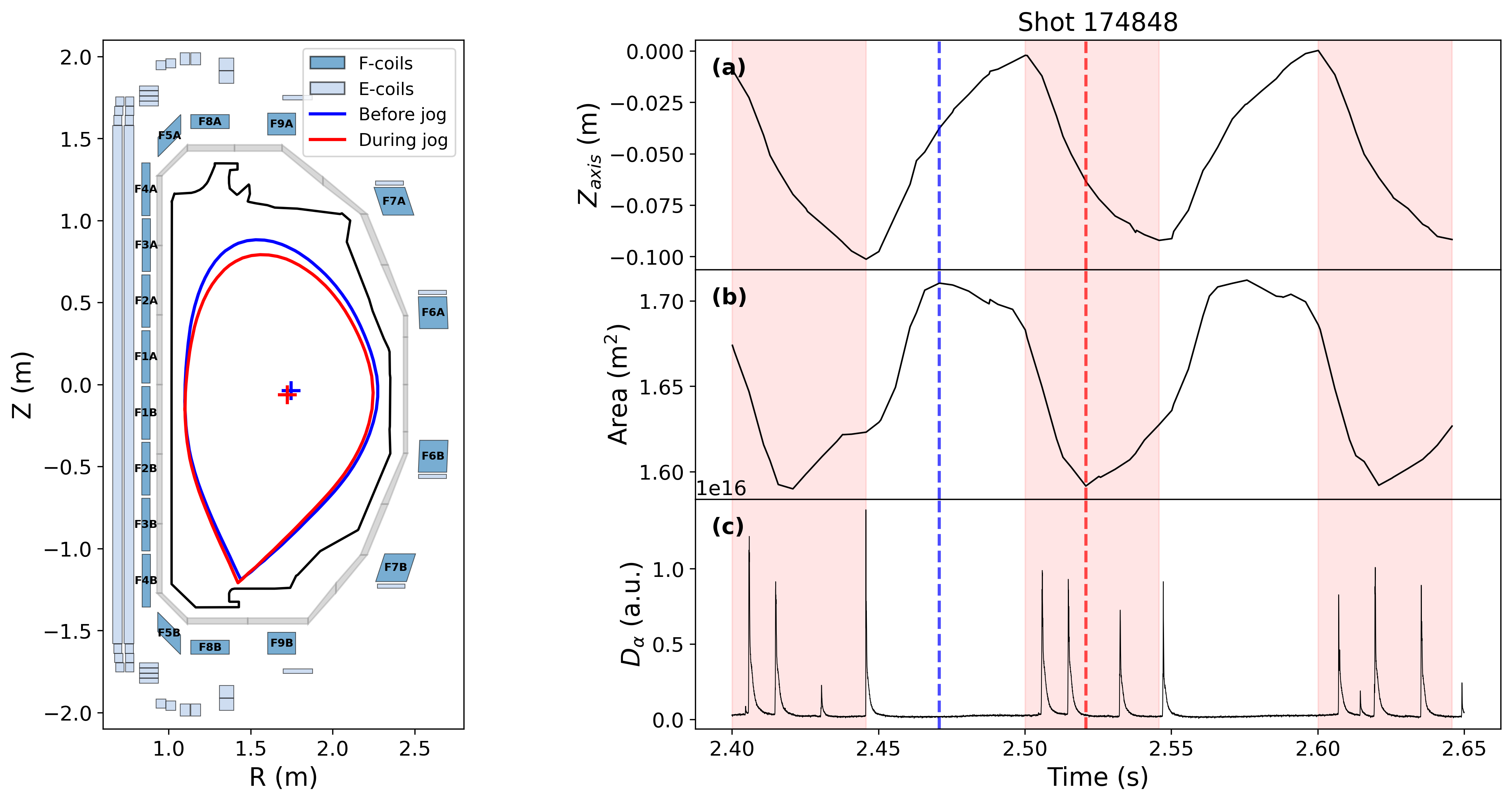}
    \caption{Left: Cross-section of the DIII-D tokamak showing the F-coil and E-coil configuration, the first wall, and the vacuum vessel. The plasma boundary from EFIT03 is shown before the jog (blue, $t = 2470$~ms) and during the jog (green, $t = 2521$~ms) for shot 174848. Right: Time traces of (a) vertical position of the magnetic axis $Z_{\mathrm{axis}}$, (b) plasma cross-section area, and (c) divertor $D_\alpha$ signal showing ELM spikes. Dashed lines indicate the two equilibrium time slices shown on the left. Note that the plasma is compressed and the cross-section area decreases by roughly 5\% during the downward motion.}
    \label{fig:coil_config}
\end{figure}

\section{Introduction}
High confinement or H-mode plasmas have characteristically good energy and particle confinement \cite{wagner_regime_1982} due to the formation of an edge transport barrier (ETB). In H-mode plasmas, there is a region near the edge with a sharp pressure gradient called the pedestal. A pedestal with larger height raises the core temperature and density of the plasma. The high density and high temperature achieved in H-mode make it an optimal operation scenario for many future fusion devices. However, the pedestal is susceptible to MHD instabilities as a result of a large bootstrap current and a steep pressure gradient. As a result of such instabilities called edge localized modes (ELMs), the ETB is degraded, releasing particles and energy from the bulk plasma  \cite{zohm_edge_1996}. It is commonly accepted that the ELM cycle depends on peeling-ballooning (P-B) modes \cite{Connor_1998, snyder_edge_2002}. When the pedestal pressure gradient or pedestal current density exceeds the P-B stability limit, there is an ELM crash relaxing the sharp pedestal pressure gradient and large bootstrap current. Naturally, the pressure gradient and current grow to pre-ELM levels on the confinement timescale and current diffusion timescale, respectively \cite{nelson_microtearing_2021}. When the plasma reaches the stability limit, there is an ELM and the cycle repeats.

Due to an ELM, up to around 10\% of the plasma stored energy can be lost in milliseconds. The resulting high transient peak heat fluxes on the plasma-facing components (PFCs) are typically not damaging for current experimental devices, but they can be detrimental for future devices. Calculations for future devices such as ITER and SPARC predict that unmitigated ELMs will result in significant melting damage to the divertor and the first wall \cite{loarte_progress_2014, Kuang_2020}. Meanwhile, ELMs can remove harmful impurities from the bulk plasma. Impurities, specifically ions of the PFC material, enter the plasma from erosion of the PFCs and can build up uncontrollably inside the ETB. Tungsten is the favored PFC material for burning plasma operations due to its high melting point and low sputtering yield \cite{Philipps_2011}.  However, tungsten is a high-Z element, i.e., it has a large atomic number Z=74, and high radiation losses from its ions limit performance and increase the risk of radiative collapse. Impurities in the bulk plasma can be expelled or pumped out across the ETB when there is an ELM. Pacing ELMs and controlling the rate of particle pump-out can limit the buildup of impurities.

Several techniques have been developed to address the transient heat loads from ELMs. Application of resonant magnetic perturbations (RMPs) by external coils can mitigate ELMs or suppress ELMs entirely \cite{evans_suppression_2005}. ITER is designed with internal coils to apply RMPs. It is also possible to operate in regimes with small or no ELMs. ELMs with large energy losses triggered at high pedestal density and high pressure gradients by coupled P-B modes are categorized as type-I ELMs. Grassy ELMs \cite{oyama_grassy_2005} and type-III ELMs have smaller energy losses of $\lesssim$ 1\% at higher frequency and are understood to be triggered by ballooning modes at high poloidal beta and the current limit, respectively. There are also ELM-free regimes such as the QH-mode \cite{burrell_qh_2002} or I-mode \cite{whyte_i-mode_2010}. Operations in small ELM or ELM-free regimes can avoid large type-I ELMs. Other techniques include pellet injection, edge electron cyclotron heating (ECH) \cite{rossel_edge-localized_2012}, and vertical jogs \cite{de_la_luna_understanding_2016, artola_non-linear_2018}. Pellet injectors that launch frozen pieces of deuterium are also being designed for ITER.

While ELM-free regimes prevent transient heat loads from ELMs, alternative means of impurity pump-out need to be developed to compensate for the lack of ELM-induced impurity transport. ELM triggering by pellet injection or vertical oscillations allows ELMs to be paced at a desired frequency to reduce the heat load while pumping out impurities. ELM pacing has been demonstrated on multiple machines through fuel or lithium pellet injection \cite{lang_elm_2004, bortolon_high_2016}, non-axisymmetric magnetic field modulation \cite{solomon_elm_2012}, and by fast vertical oscillations or jogging of the plasma \cite{de_la_luna_understanding_2016, gerhardt_first_2010}. In particular, jogging does not require additional hardware specifically to inject pellets or apply external fields.

Fast vertical oscillations of the plasma called jogging have long been investigated as a method to trigger ELMs. First reported on TCV \cite{degeling_magnetic_2003, cruz_control_2018}, ELM triggering by vertical plasma displacement has been demonstrated at AUG \cite{lang_frequency_2004}, NSTX \cite{gerhardt_first_2010}, KSTAR \cite{kim_elm_2012}, JET \cite{de_la_luna_magnetic_2009, de_la_luna_understanding_2016}, and HL-2A \cite{wu_experiment_2017,wu_study_2018}. ELMs have been triggered at higher than natural frequencies with smaller energy losses. The mechanism of ELM triggering is understood to be due to the perturbation of the edge current above a critical value, which destabilizes the peeling instability \cite{artola_non-linear_2018}. In this case, the induction of toroidal current in the pedestal is a result of plasma motion through an inhomogeneous magnetic field. Figure~\ref{fig:coil_config} shows ELM triggering by jogging at DIII-D.

The purpose of this work is to report the observation of ELM triggering via vertical jogs at DIII-D and explain the underlying physics processes. Dedicated experiments have been conducted in DIII-D to prove the effectiveness of ELM pacing via jogging and understand its mechanism. The experimental efforts have been accompanied by simple modeling to help in the analysis and interpretation of the experimental data. Jogging has successfully triggered smaller ELMs at higher frequency with reduced heat loads to the divertor. ELMs have been triggered favorably when the plasma in lower single null (LSN) configuration is moving toward the X-point. 

\begin{figure}[t]
    \centering
    \includegraphics[width=0.68\linewidth, height=0.68\textheight, keepaspectratio]{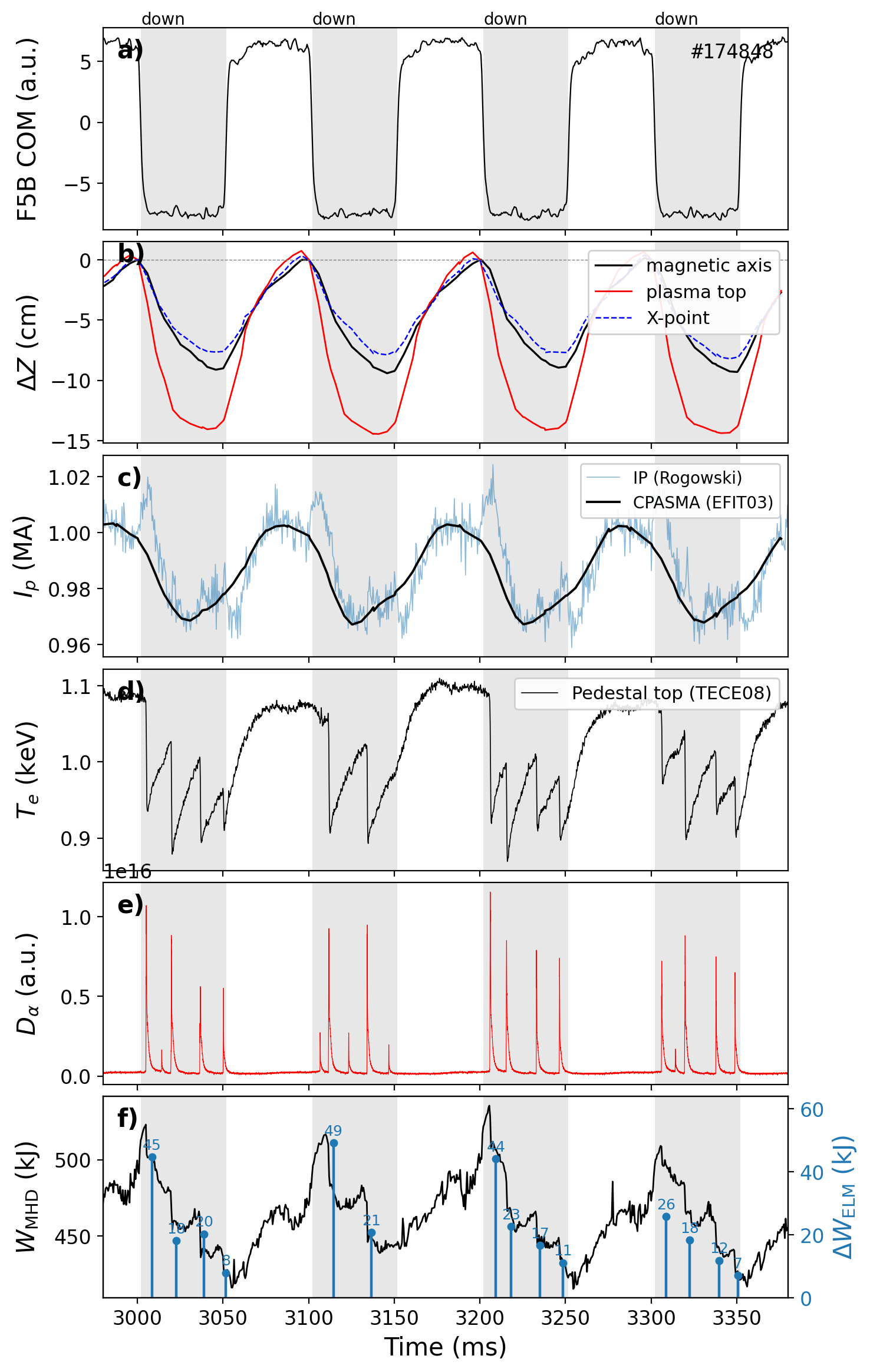}
    \caption{Plasma parameters during a series of downward vertical jogs in DIII-D shot 174848: (a) The F5B coil command signal, (b) $\Delta Z$, the relative vertical displacement (with respect to the pre-jog position) of the magnetic axis, the plasma top, and the X-point, (c) the plasma current $I_p$, (d) electron temperature from electron cyclotron emission (ECE) at the pedestal top, (e) the divertor $D_\alpha$, and (f) the stored energy $W_\mathrm{MHD}$ (black curve, left axis) together with the per-ELM energy loss $\Delta W$ (blue stems, right axis). Note that the movement of the X-point is smaller than the movement of the plasma top.}
    \label{fig:pcs_command}
\end{figure}

The remainder of this paper is organized as follows. The setup and results of the experiments are presented in Section~\ref{sec:experiments}. In Section~\ref{sec:result}, the mechanism of ELM triggering is explained and the results of stability analysis based on a toy model are presented. The experimental and modeled results are discussed in Section~\ref{sec:discussion}. Finally, Section~\ref{sec:conclusion} concludes this work.

\section{Experiments and Observations}
\label{sec:experiments}
\subsection{Vertical Jogs at DIII-D}
For the DIII-D discharges discussed in this work, vertical jogs are pre-programmed by artificially sending saturation level commands to each of the F-coils, the poloidal field coils. The saturation interval---the duration for which the coil commands are held at their saturation level with feedback control disabled---lasts a few milliseconds ($\lesssim 5$~ms). During this time, both vertical position and shape controllers are temporarily lost. The control system regains control of the plasma position and shape before a collision with the machine vessel walls. When reactivated, the control system brings the plasma back to the pre-jog state. During jogging, the movement of the X-point is lagging and more limited compared to the top of the plasma. This leads to the compression of the plasma; as shown in Figure~\ref{fig:coil_config}, the cross-section area of the plasma rapidly decreases by roughly 5\% during the downward motion. Throughout this entire cycle, the control system attempts to keep the plasma current constant by modifying the voltage ramp rate of E-coils, the central solenoid at DIII-D. Figure~\ref{fig:pcs_command} shows the F-coil command signals, displacement of the plasma, and subsequently triggered ELMs.

\begin{figure}
    \centering
    \includegraphics[width=\linewidth, height=0.77\textheight, keepaspectratio]{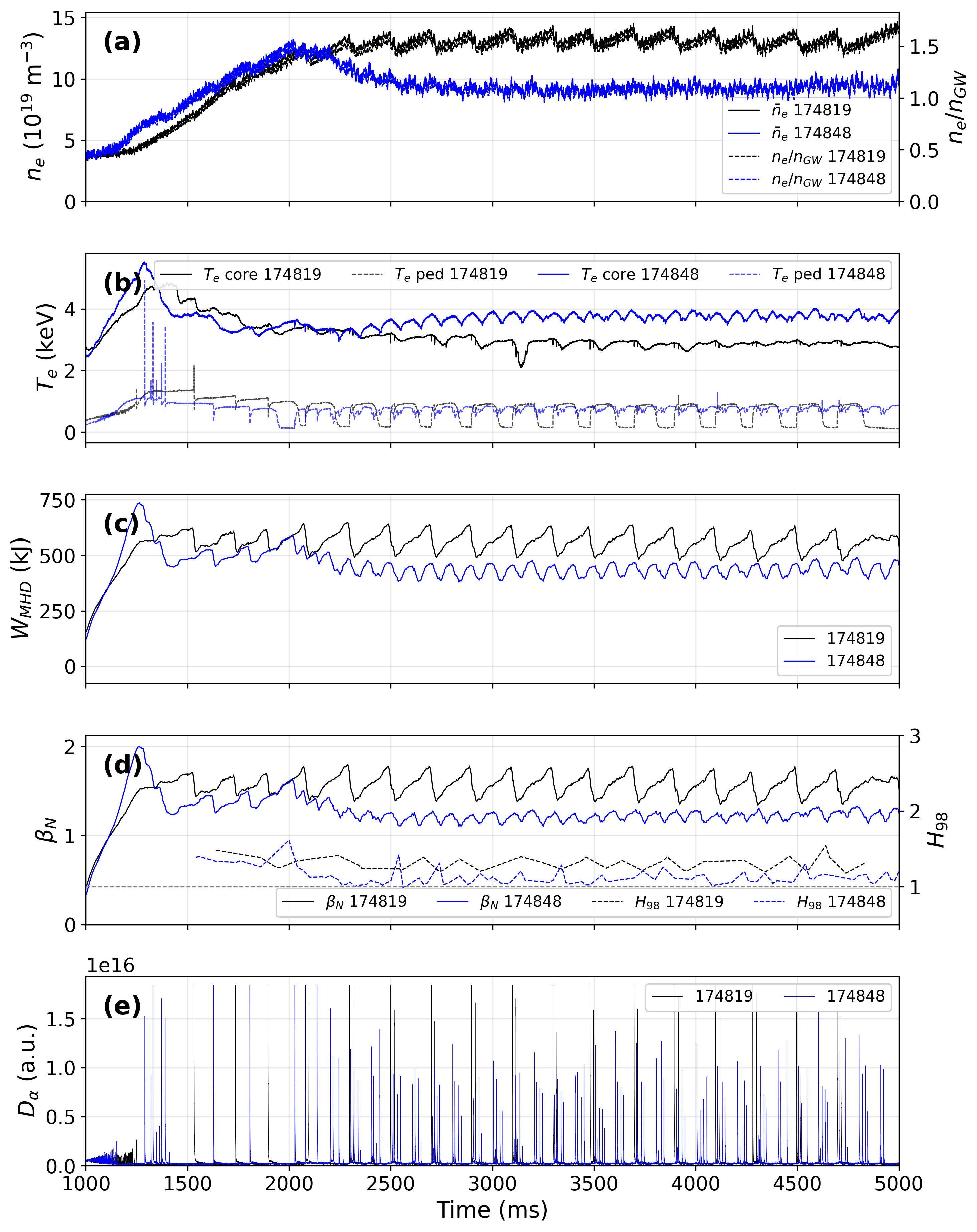}
    \caption{Time histories of key plasma parameters for shots 174819 (black) and 174848 (blue): (a) Line-averaged electron density $\bar{n}_e$ (solid) and Greenwald fraction $n_e/n_{GW}$ (dashed), (b) Electron temperature from ECE measurements at the core (solid) and pedestal-top (dashed), (c) Stored energy $W_\mathrm{MHD}$, (d) Normalized beta $\beta_N$ (solid) and confinement factor $H_{98,y2}$ (dashed), (e) Divertor $D_\alpha$. The jogging discharge 174848 exhibits more frequent, smaller ELMs than the reference 174819.}
    \label{fig:timetrace}
\end{figure}

All of the discharges discussed in this work had an LSN configuration with a flat-top plasma current of $I_p\sim$1~MA and a toroidal field strength of $B_t\sim$2.1~T. The main discharge analyzed in this work is DIII-D shot 174848. This discharge clearly demonstrates the effectiveness of ELM control via vertical jogs: smaller and more frequent ELMs are triggered during the jogging phase, leading to reduced heat loads to the divertor and reduced impurity content.

\begin{figure}
    \centering
    \includegraphics[width = 3.5 in, keepaspectratio]{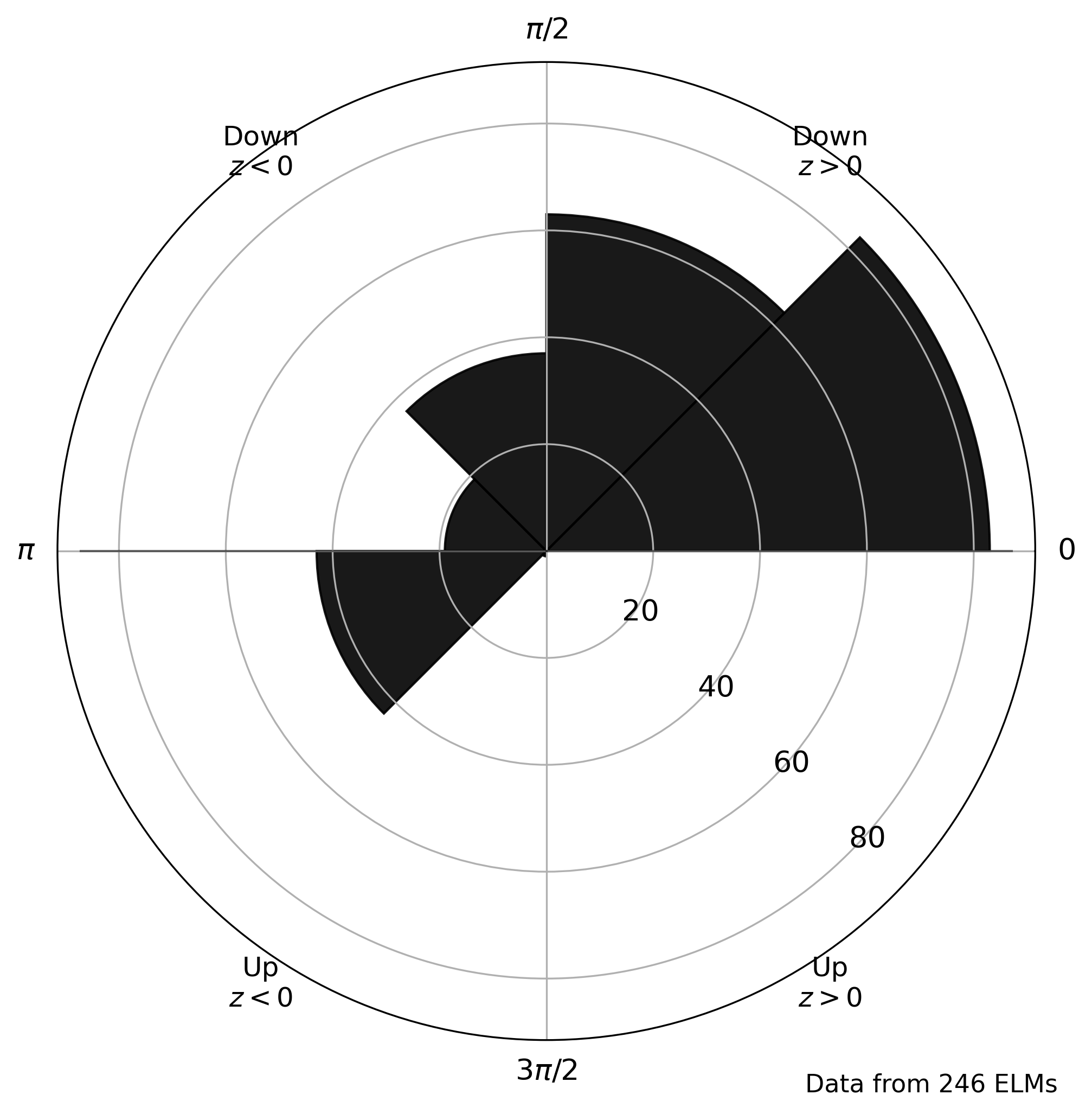}
    \caption{Polar histogram of ELM-jogging phase for shot 174848. The shaded region indicates the phase at which the ELMs occurred and the cumulative count of ELMs.}
    \label{fig:Polar_ELM}
\end{figure}

\subsection{Characterization of ELM Triggering}
The effect of vertical jogs on key plasma parameters is illustrated in Figure~\ref{fig:timetrace}, which compares time traces of key parameters for shot 174848 and the reference shot 174819 without jogs. Before the onset of jogging, the two discharges exhibit similar line-averaged electron density $\bar{n}_e$, core and pedestal electron temperatures, stored energy $W_\mathrm{MHD}$, and normalized beta $\beta_N$. Following the onset of jogging, the stored energy $W_\mathrm{MHD}$ and $\beta_N$ are reduced on shot 174848 relative to the reference, by of order 20--30\%. In 174848, the energy losses and temperature drops from individual ELMs are also visibly smaller than in the reference shot. The confinement factor $H_{98,y2}$ nevertheless remains close to unity throughout both discharges.

In order to quantify the relationship between the vertical jogs and the ELMs, an ELM-jog phase is defined, similar to that defined in prior work on NSTX \cite{gerhardt_first_2010}. The jog period is defined to begin when the magnetic axis is at its mean vertical position and moving downwards. Each ELM is then assigned a phase based on its timing within this period: an ELM at the start of the period has zero phase, and the phase increases linearly to $2\pi$ at the end of the period:

\begin{equation}
    \phi_{ELM-jog} = 2\pi\frac{t_{ELM} - t_{jog,start}}{t_{jog,end}-t_{jog,start}}
\end{equation}

\noindent where $t_{ELM}$ is the time of the ELM, and $t_{jog,start}$ and $t_{jog,end}$ are the start and end times of the jog period, respectively. The ELM-jog phases for each ELM during the jogging duration of shot 174848 are shown in Figure \ref{fig:Polar_ELM}. From the figure, it can be seen that a majority of the ELM-jog phases lie in the range $0\leq\phi_{ELM-jog}\leq \pi/2$, which corresponds to the ELM occurring while the plasma is moving vertically downwards, with the rest of the ELMs occurring shortly after the plasma reaches the trough. This demonstrates a clear correlation between the vertical jogs and the triggered ELMs. It also should be noted that every vertical jog in shot 174848 triggered ELMs. Multiple ELMs are triggered per jog, a phenomenon which has been previously observed on KSTAR \cite{kim_elm_2012} and sporadically at AUG \cite{lang_frequency_2004}. These compound ELMs consist of a primary ELM triggered during the downward motion followed by secondary ELMs during the recovery phase. The onset of multiple ELMs per jog suggests that the induced edge current by jogging is sufficiently strong to trigger multiple instabilities before the edge current relaxes as the pedestal recovers between consecutive ELMs within a single jog cycle.

\begin{figure}[t]
    \centering
    \includegraphics[width = 3.5 in, keepaspectratio]{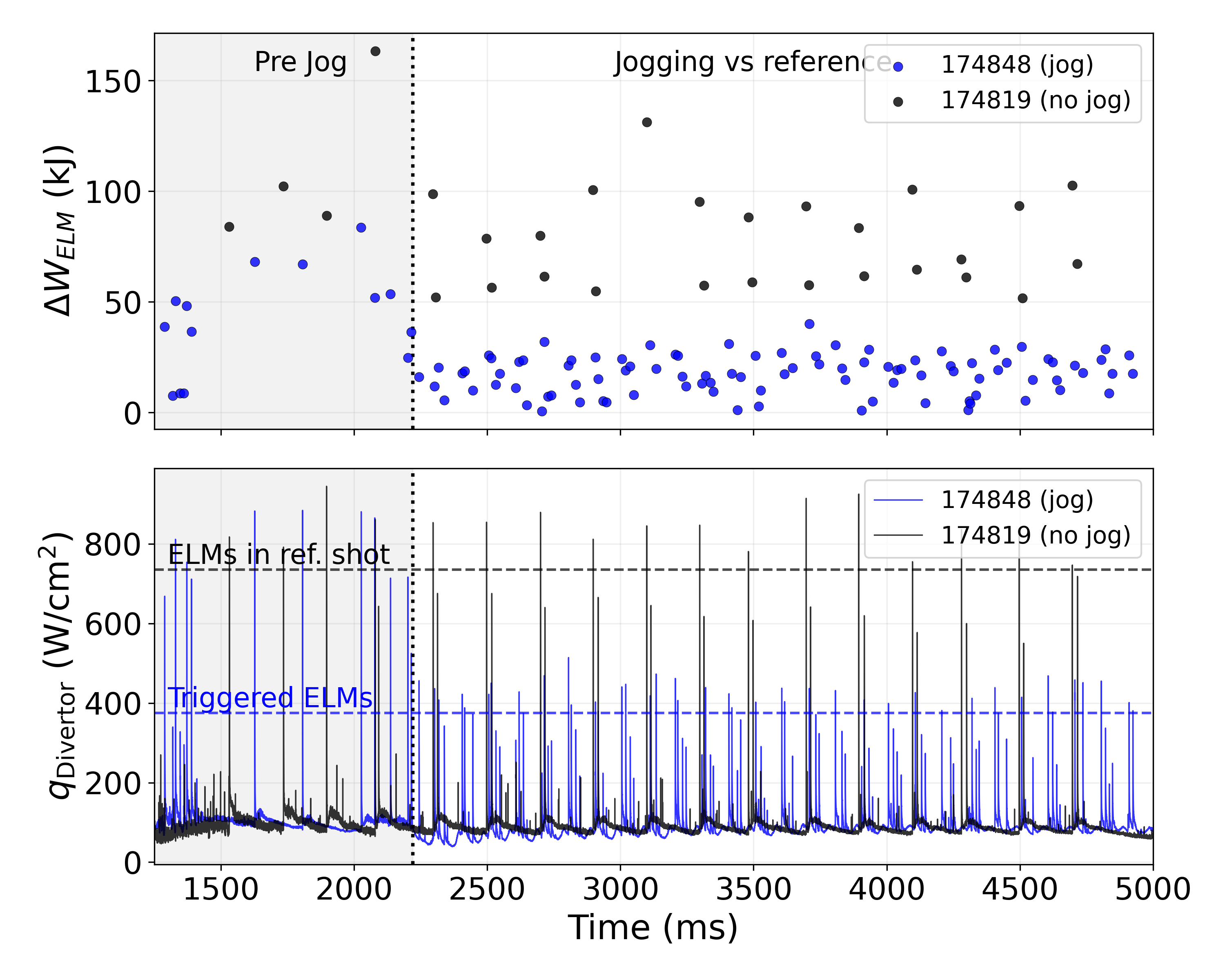}
    \caption{Time trace of change in stored energy per ELM (above) peak heat flux to lower outer divertor as measured by IRTV (below) for shot 174848 (blue) and 174819 (black). After $\sim$2.0s, jogging begins in shot 174848.}
    \label{fig:heat_flux}
\end{figure}

\subsection{Effects of ELM pacing}
A primary motivation for ELM pacing is the reduction of transient heat loads to the divertor. In Figure~\ref{fig:heat_flux}, the peak heat flux to the outer divertor, as measured by infrared thermography (IRTV), of shot 174848 and reference shot 174819 are compared. Before the onset of jogging, both discharges exhibit similar peak heat flux levels due to natural ELMs. After jogging begins, the peak heat flux spikes on shot 174848 are reduced by approximately a factor of 2 compared to the natural ELMs on the reference shot. This reduction is a direct consequence of a reduction of individual ELM energy losses.

\begin{figure}
    \centering
    \includegraphics[width = 3.5 in, keepaspectratio]{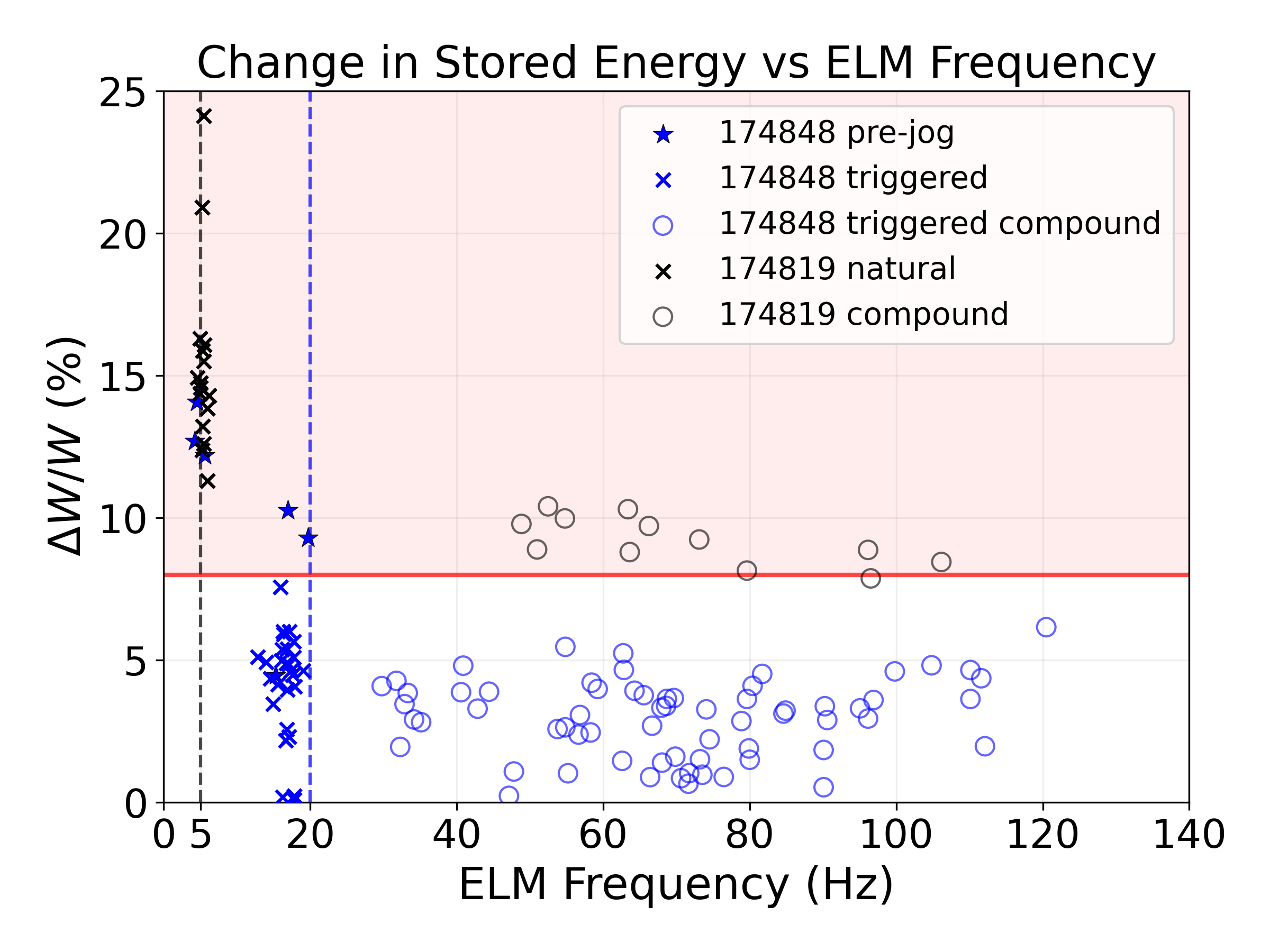}
    \caption{Change in plasma stored energy after ELM (normalized to plasma stored energy immediately before the ELM) plotted against inverse of time since last ELM.}
    \label{fig:WMHD}
\end{figure}

Figure~\ref{fig:WMHD} shows the fractional change in plasma stored energy $\Delta W/W$ after each ELM against the ELM frequency or the inverse of time since the previous ELM. ELMs that are triggered at higher than the natural frequency exhibit smaller energy loss fractions. While up to $\sim$25\% of the energy is lost from the natural ELMs, energy loss by triggered ELMs does not exceed $\sim$8\% and can be significantly smaller. The maximum energy loss fraction from the triggered ELMs does appear to decrease at higher frequencies, although the measurement uncertainty introduced by jogging and ELMs is difficult to ignore.

\begin{figure}[t]
    \centering
    \includegraphics[width = 3.5 in, keepaspectratio]{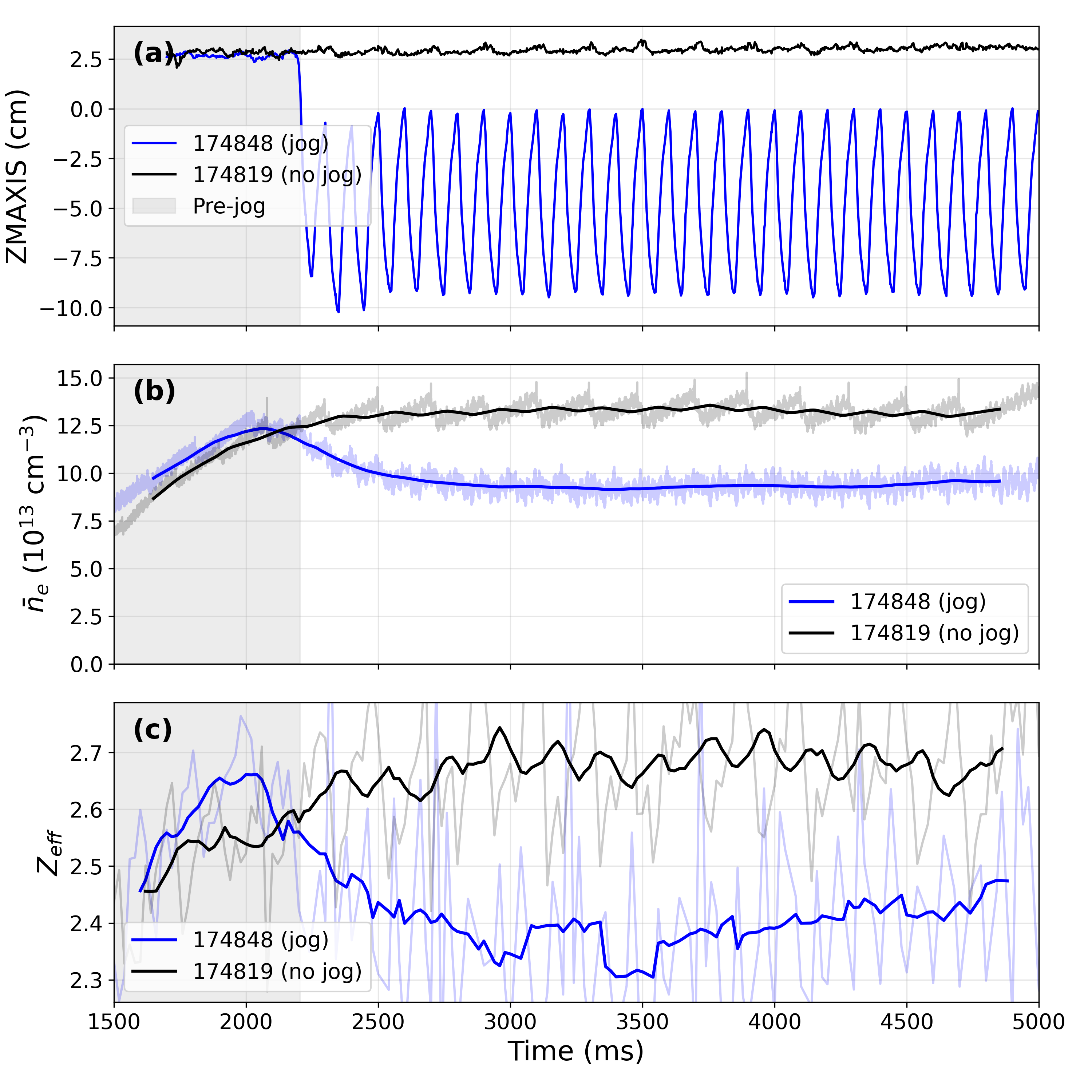}
    \caption{(a) Magnetic axis vertical position, (b) core line-averaged electron density, and (c) line-averaged $Z_\mathrm{eff}$ from visible bremsstrahlung for shot 174848 (blue) and reference shot 174819 (black). Note that the lower $Z_\mathrm{eff}$ on 174848 is not a consequence of the reduced electron density: $Z_\mathrm{eff}$ can be derived from the visible bremsstrahlung emission as $Z_\mathrm{eff} \sim V_B \sqrt{T_e} / n_e^2$, a lower $n_e$ with unchanged impurity content would actually produce a higher $Z_\mathrm{eff}$, not a lower one. The reduction in $Z_\mathrm{eff}$ therefore implies a decrease in impurity content as a result of jogging.}
    \label{fig:ZEFF}
\end{figure}

In addition to reduced heat loads, the vertical jogs lead to a measurable reduction in impurity content. Figure~\ref{fig:ZEFF} shows the magnetic axis vertical position, core line-averaged electron density, and line-averaged $Z_\mathrm{eff}$ between the jogging and reference discharges. Following the onset of jogging, the line-averaged $Z_\mathrm{eff}$ on shot 174848 is reduced relative to the reference shot throughout the jogging phase. This reduction is not an artifact of a reduction in electron density; $Z_\mathrm{eff}$ is derived from visible bremsstrahlung emission as $Z_\mathrm{eff} \sim V_B \sqrt{T_e}/n_e^2$, so a lower $n_e$ would result in an increase in $Z_\mathrm{eff}$. The reduction of $Z_\mathrm{eff}$ therefore implies a decrease in impurity density as a consequence of jogging. This can be attributed to a combination of increased impurity pump-out by ELMs and reduced impurity influx from the PFCs due to smaller ELMs thus reduced sputtering.

\section{Modeling of ELM triggering by vertical oscillations} 
\label{sec:result}

\subsection{Mechanism of edge current induction}
\begin{figure}[t]
    \centering
    \includegraphics[width = 3 in, keepaspectratio]{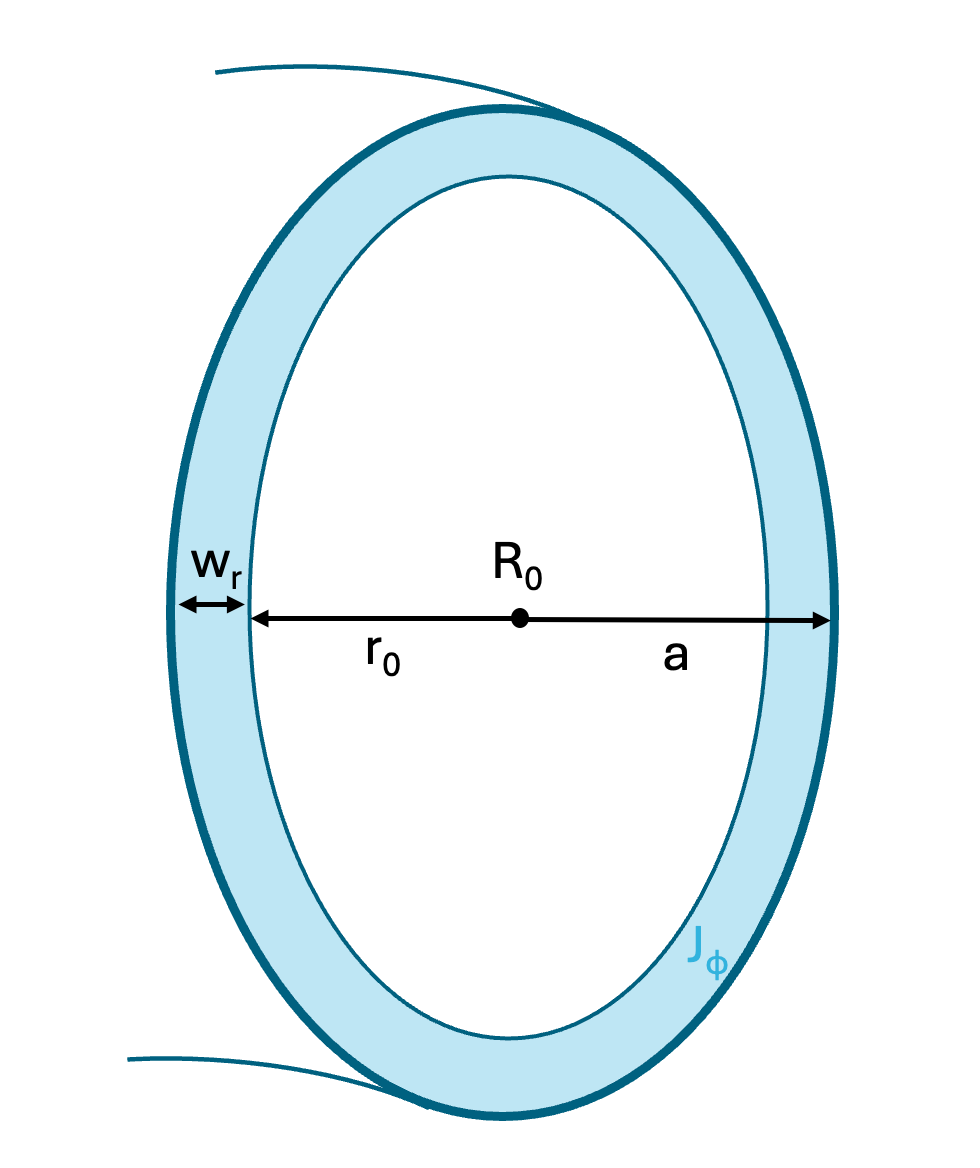}
    \caption{Schematic cylindrical model used for the current induction analytical calculation.}
    \label{fig:cylindrical_model}
\end{figure}

ELM triggering has been attributed to an increase of the edge toroidal current driving the plasma into an unstable MHD P-B regime. Simulations of ITER and JET plasmas suggest that vertical jogs indeed induce an edge current in the edge region \cite{artola_non-linear_2018}. Equations for edge current induction by vertical jogs are derived from the MHD equation for the poloidal flux by Artola et al., for the simplified cylindrical plasma geometry shown in Figure~\ref{fig:cylindrical_model}:

\begin{equation}
    \delta I_{\phi}^{w_r} = \frac{4\pi}{\mu_0 R_0} [\delta \psi_{ext}(a) - B_{\theta}(r_0)R_0\delta w_r - \eta J_{\phi} \delta t]
    \label{eq:edge_current}
\end{equation}

with the change in current density $\delta J_{\phi}$ given as

\begin{equation}
    \delta J_{\phi} = \frac{1}{2\pi r_0w_r} \left( \delta I_{\phi}^{w_r} - I_{\phi}^{w_r} \frac{\delta w_r}{w_r} \right)
\end{equation}

where $r_0$ is the radius to the plasma edge, $w_r$ is the width of the edge region---the resistive skin layer just inside the separatrix, of order the electromagnetic skin depth and corresponding to the outer pedestal region ($\hat\psi \gtrsim 0.92$), the range over which the ELITE edge current density is evaluated---$a$ is the minor radius, $\psi_{ext}$ is the external poloidal flux contribution---the poloidal flux produced by currents flowing outside the plasma, such as those in the poloidal field coils and the conducting vessel structures, as distinct from the flux generated by the plasma current itself---and $\eta$ is the resistivity of the plasma. Contributions to the total edge current $I_{\phi}^{w_r}$ are from
\begin{itemize}
    \item Local changes in the external flux in time $\delta \psi_{ext}$, for example those produced by time-varying currents in the external poloidal field coils
    \item Plasma motion through an inhomogeneous magnetic field $\delta \mathbf{r}\cdot \nabla \psi_{ext}$
    \item Plasma compression or expansion (through $\delta w_r$)
\end{itemize}

The edge width can be approximated as the electromagnetic skin depth $w_r\sim \sqrt{\eta/(\pi\mu_0f)}$, where $f$ is the vertical-oscillation frequency, set for each discharge from the measured jog transit time between successive vertical extrema, $f = 1/(2\tau_{\mathrm{stroke}})$. The global resistive diffusion time $\tau_R \sim \mu_0 a^2 / \eta$ is of order $10-10^2$~s (temperature at pedestal top $T_e\sim500$~eV to the core $T_e\sim$~few keV)---far longer than the downstroke period $\tau_{\mathrm{stroke}} \sim 50$~ms. As a result, the induced edge current cannot diffuse radially and instead remains confined to the narrow edge layer of width $w_r \ll a$. The resistive diffusion time across this layer, $\mu_0 w_r^2/\eta = 1/(\pi f)$, is by construction comparable to the oscillation period, ensuring that the induced current remains localized at the plasma edge where it can affect P-B stability.

The competing contributions of compression and motion through the inhomogeneous magnetic field determine the direction and magnitude of the induced edge current. As the plasma moves toward the lower X-point, it approaches the like-signed coil current that forms the X-point, and by Lenz's law the resistive edge responds with a screening current of the \emph{opposite} sign---directed opposite to the plasma current. This is the $\delta\psi_{ext}(a)$ (and $\delta\mathbf{r}\cdot\nabla\psi_{ext}$) contribution in Eq.~(\ref{eq:edge_current}), which is negative on the downstroke ($\delta I < 0$), following Artola \emph{et al.}~\cite{artola_non-linear_2018}. Simultaneously, as the plasma is compressed the edge width $w_r$ narrows ($\delta w_r < 0$), so the compression term $\delta I = - B_{\theta}(r_0) R_0 \delta w_r > 0$ drives a current in the \emph{same} direction as the plasma current. The respective contributions to the edge plasma currently compete; during the downward jog in 174848 the compression term dominates, so the \emph{net} induced edge current is positive, providing the peeling drive. During an upward jog the signs reverse.

\subsection{Application of the edge current evolution toy model}
Based on the edge current induction mechanism, a toy model of the vertical oscillation has been created and implemented with TokaMaker \cite{Hansen2024, hansen_oft_2026}. First, an equilibrium from kinetic EFIT gives the initial pressure and current profiles. At each time step, the TokaMaker code solves the free-boundary equilibrium equation based on the coil current values from DIII-D experiments. The imposed boundary shapes are from EFIT03 reconstructions. EFIT03 is a kinetic equilibrium reconstruction using magnetic measurements, motional Stark effect (MSE) data, and kinetic profile information from Thomson scattering and charge exchange recombination spectroscopy with accurate current density profiles, particularly in the pedestal region. The change in edge current and edge current density are calculated based on the equations above. The cumulative compression-induced edge current density is incorporated as a non-inductive contribution to the $FF'$ profile, where $F = R B_\phi$ is the toroidal field function and $FF' = F\,\mathrm{d}F/\mathrm{d}\psi$ is its derivative with respect to the poloidal flux. The spatial profile is modeled as a Gaussian localized at the skin depth:

\begin{equation}
    J_{\mathrm{NI}}(\hat\psi) = J_{\mathrm{edge}} \exp\left[
    -\frac{(\hat\psi - \hat\psi_c)^2}{2\,w_\psi^2}\right],
    \label{eq:gaussian_NI}
\end{equation}

where $\hat\psi_c = 1 - w_\psi$ is the normalized flux at the Gaussian center, $w_\psi = w_r / a_{\mathrm{eff}}$ maps the physical skin depth to flux space, and $a_{\mathrm{eff}} = \sqrt{A/\pi}$ is the effective minor radius. This profile is converted to the non-inductive $FF'$ contribution via

\begin{equation}
    FF'_{\mathrm{NI}} = \frac{\mu_0 \, J_{\mathrm{NI}}}{\langle 1/R \rangle},
\end{equation}

where $\langle 1/R \rangle$ is the flux-surface-averaged inverse major radius from the previous equilibrium solution. The evolution of the edge width $w_r$ and thus the temperature in the edge resistive layer is key to this toy model. Obtaining accurate temperature measurements of the edge region during jogging is difficult as the diagnostic measuring points are constantly changing due to the rapid movement of the plasma. The rapid downward movement and the subsequent compression occur over the $\sim$50~ms downstroke, still shorter than the energy confinement time $\tau_E$. Thus, it can reasonably be assumed that the plasma is adiabatic. The edge temperature scales with the plasma volume as

\begin{equation}
    T_{\mathrm{edge}} \propto V^{-(\gamma - 1)},
    \label{eq:adiabatic_T}
\end{equation}

where $V(t) = 2\pi R_0 A(t)$ is the plasma volume and $\gamma = 5/3$ is the adiabatic index. The cross-section area $A(t)$ is obtained from the EFIT03 boundary. The axis pressure target is also updated adiabatically at each step: $p_{\mathrm{ax}}(t) = p_{\mathrm{ax,ref}} (V_{\mathrm{ref}}/V(t))^\gamma$. The edge resistivity is calculated from the simplified Spitzer formula, $\eta \propto T_{\mathrm{edge}}^{-3/2}$, normalized to a reference pedestal-top temperature $T_{\mathrm{edge,ref}} \approx 500$~eV taken from the kinetic-EFIT profiles, and the electromagnetic skin depth scales directly with the plasma volume:

\begin{equation}
    w_r \propto V^{3(\gamma-1)/4}.
    \label{eq:skin_depth_V}
\end{equation}

As the plasma compresses ($V$ decreases), $T_{\mathrm{edge}}$ increases, $\eta$ decreases, and $w_r$ narrows---concentrating the induced edge current into a thinner edge layer. This relationship is consistent with the experimental observations in Figure~\ref{fig:pcs_command}, where the plasma cross-section is compressed during the rapid downward movement followed by ELMs. The \emph{motion} term comes from the vertical displacement of the magnetic axis (the $\delta\mathbf{r}\cdot\nabla\psi_{ext}$ screening term), while the \emph{compression} term comes from the decrease in plasma cross-sectional area (the $\delta w_r$ term).

\begin{figure}[htbp]
\centering
\includegraphics[width=0.78\linewidth, height=0.8\textheight, keepaspectratio]{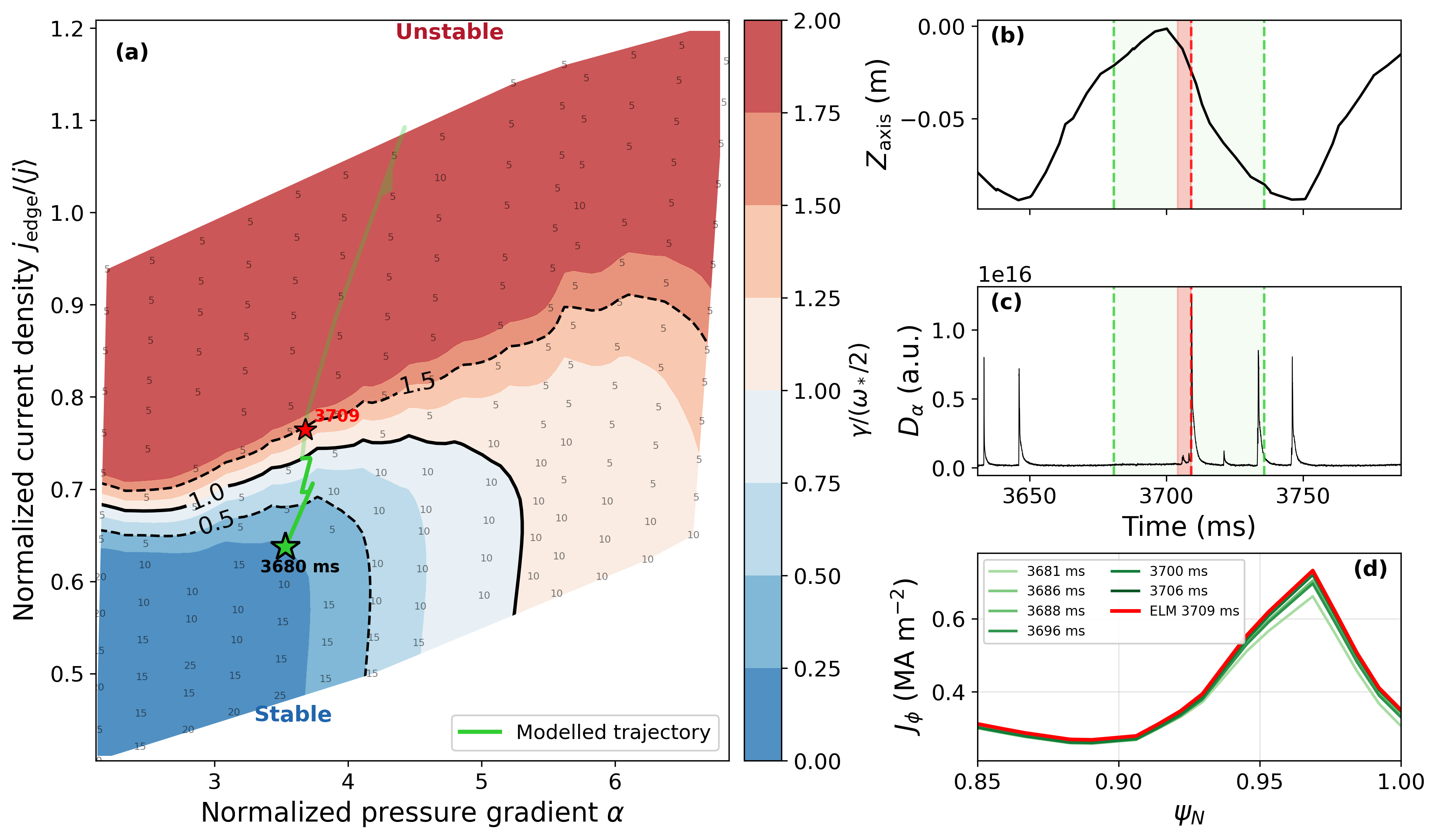}
    \caption{(a) P-B stability diagram for DIII-D shot 174848 at $t = 3680$~ms from ELITE: the contour map shows the normalized growth rate $\gamma/(\omega_*/2)$ versus the normalized pressure gradient $\alpha$ and normalized edge current density $j_\mathrm{edge}/\langle j \rangle$. The green curve is the modeled evolution of the edge operating point, with the green star the reference equilibrium time ($t = 3680$~ms) and the red star the ELM time. (b) Magnetic-axis position $Z_\mathrm{axis}$; the green band spans the simulation window and the red band marks the ELM onset. (c) The $D_\alpha$ signal over the same window. (d) Edge toroidal current-density profile $J_\phi(\psi_N)$ at successive time points along the trajectory up to the ELM; the peak near $\psi_N \approx 0.97$ builds up as the operating point crosses the peeling boundary and moves deeper into the unstable region. Note the current density relaxation by an ELM is not modeled, thus, the modeled trajectory is in faint green after the ELM crash.
    }
    \label{fig:ELITE}
\end{figure}

\begin{figure}[t]
\centering
\includegraphics[width=0.58\linewidth, height=0.72\textheight, keepaspectratio]{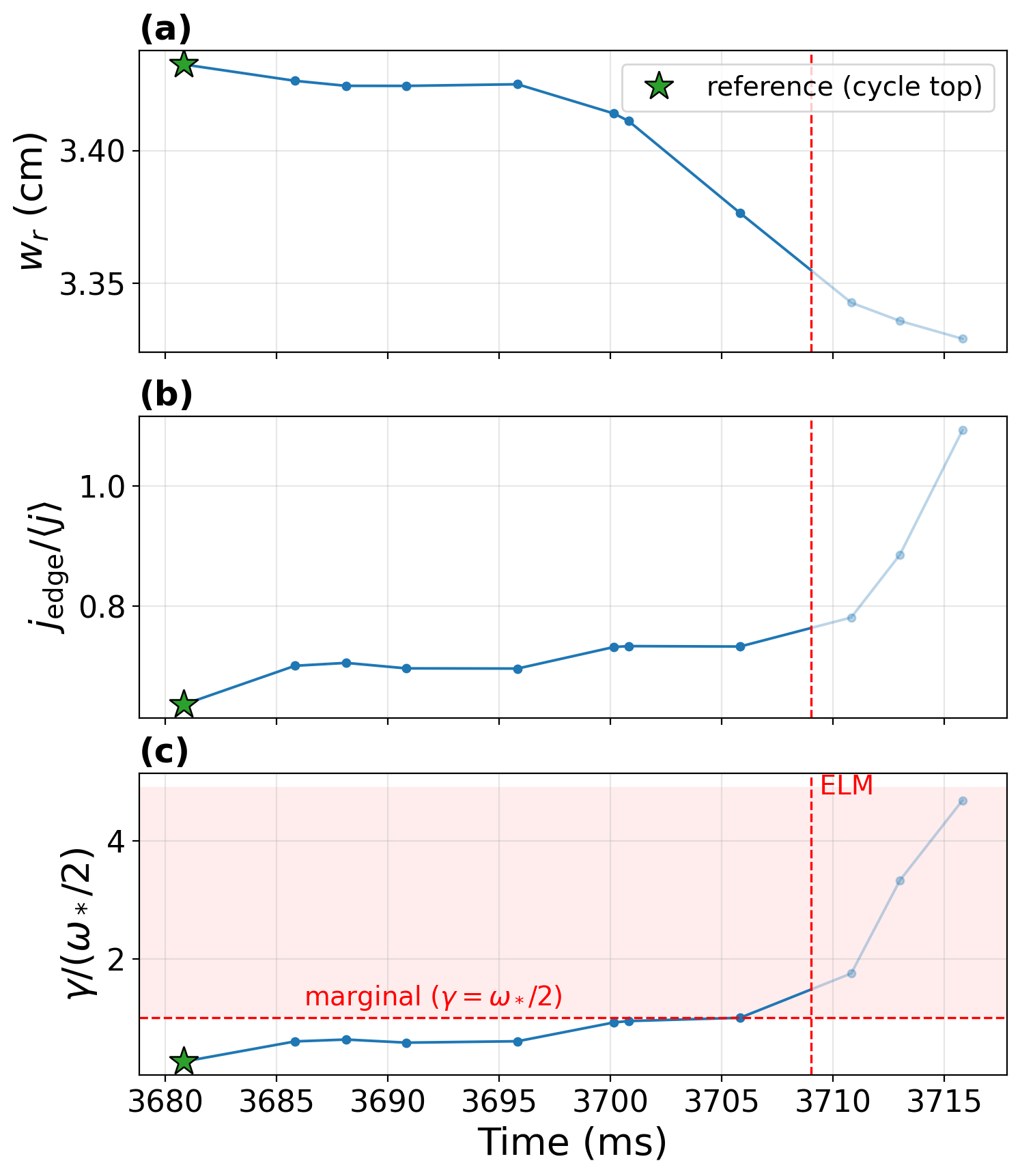}
    \caption{Time evolution over the modeled $t=3680$~ms downstroke of shot 174848, showing the three quantities that govern the peeling stability versus time: (a) the resistive skin depth $w_r$, (b) the normalized edge current density $j_\mathrm{edge}/\langle j\rangle$, and (c) the normalized growth rate $\gamma/(\omega_*/2)$. The green star marks the reference equilibrium and the red dashed line the observed ELM onset. As the plasma compresses on the downstroke the skin depth narrows, concentrating the induced edge current so that $j_\mathrm{edge}/\langle j\rangle$ and the growth rate rise, carrying the operating point across the marginal-stability boundary and into the unstable region, where the ELM is triggered. Note the traces are faded after the ELM as post-ELM evolution is no longer correctly described by the model.}
    \label{fig:growthrate_3680}
\end{figure}

\begin{figure}[t]
\centering
\includegraphics[width=0.83\linewidth, height=0.70\textheight, keepaspectratio]{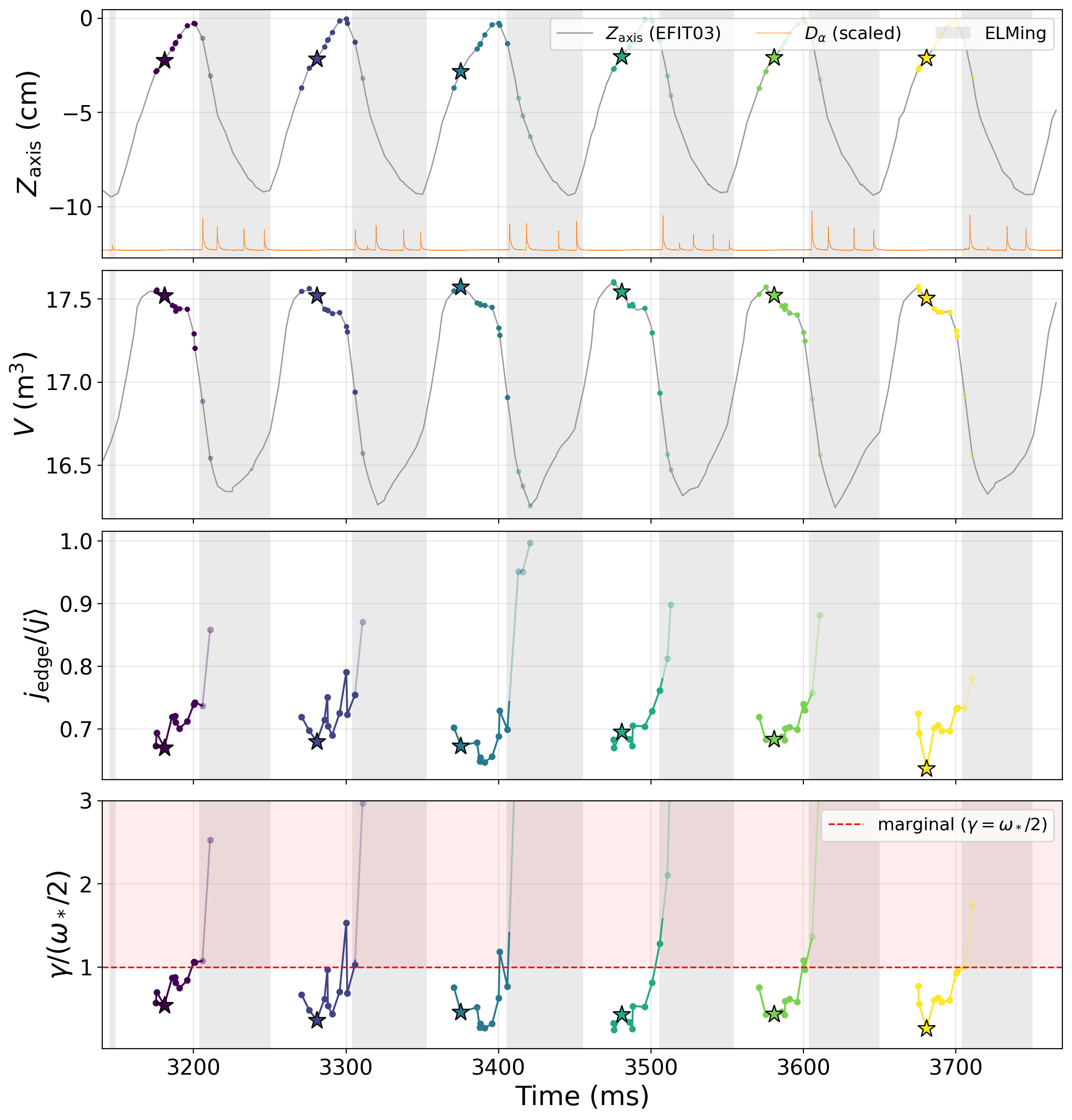}
    \caption{P-B stability evolution over multiple jog cycles of shot 174848. For each cycle, the edge current and equilibrium are obtained by integrating both backward and forward from that cycle's reference equilibrium. ELMing intervals are shaded and colored markers trace each cycle's trajectory and stars mark the reference. Key parameters are (a) Magnetic-axis position $Z_\mathrm{axis}$ (grey) and divertor $D_\alpha$ (orange), (b) Plasma volume $V$, (c) Normalized edge current density $j_\mathrm{edge}/\langle j\rangle$ along each trajectory, (d) Normalized growth rate $\gamma/(\omega_*/2)$ along each trajectory. On every one of the six jog cycles the operating point is driven across the marginal boundary within a few milliseconds of the observed ELM, demonstrating that the induced-edge-current mechanism triggers the ELM on each downstroke.}
    \label{fig:stability_timetrace}
\end{figure}

At each timestep, the measured boundary shape defines a new plasma volume, which determines the adiabatically compressed edge temperature and, through Spitzer resistivity, the edge width. The narrowing skin depth concentrates the compression-driven edge current into a thinner layer near the separatrix, whose amplitude is evolved using the equations. TokaMaker~\cite{Hansen2024} solves for a new equilibrium with the accumulated non-inductive edge current contribution and the adiabatically obtained pressure target constraint, following the edge-current induction and adiabatic-evolution formulation of Artola~\emph{et al.}~\cite{artola_non-linear_2018}. The iteration is repeated.

Temporal continuity between equilibria is enforced through the inductive constraint $\psi_{\mathrm{dt}}$, which biases the solver toward the previous step's flux distribution over the elapsed time interval $\Delta t$. Throughout this iteration, TokaMaker maintains a self-consistent equilibrium that responds to the evolving edge current. It solves the nonlinear Grad--Shafranov (GS) equation with the compression-induced edge current included as an additional source term, so that the resulting flux surfaces, current profile, and pressure gradient all adjust consistently. This coupling is essential because the edge current modifies the local magnetic shear and pressure gradient---the very quantities that determine P-B stability.

\subsection{Stability analysis during jogging}
Equilibrium evolution simulation and stability analysis using ELITE \cite{wilson_elite_2002, snyder_elms_2004} has been performed on DIII-D jogging experiments. VARYPED has been used to generate equilibrium variations by scanning the pedestal pressure gradient and edge current density around a reference kinetic equilibrium while preserving collisionality and total stored energy \cite{meneghini_omfit_2015}. These equilibrium variations are then passed to ELITE, which performs stability scans over toroidal mode numbers $n = 5$--$25$. Figure~\ref{fig:ELITE} shows the results for shot 174848. The P-B stability diagram at $t = 3680$~ms is shown on the left. The contour map displays the normalized growth rate $\gamma/(\omega_*/2)$~\cite{snyder_edge_2002} (the diamagnetic-stabilization criterion, where a mode is unstable for $\gamma > \omega_*/2$) as a function of normalized pressure gradient $\alpha$ and normalized edge current density $j_\mathrm{edge} / \langle j \rangle$.  As the diagram shows, at $t = 3680$~ms the plasma remains in the stable region far from the ballooning boundary (toward the right) and closer to the peeling boundary (toward the top). ELITE calculations of equilibria at other times with the same ELM-jog phase yield similar stability boundaries. 

The edge current evolution simulation is initialized at $t = 3680$~ms, which corresponds to a time shortly before a downward jog and $\sim 30$~ms before the first ELM. At each timestep, $\alpha$ and $j_\mathrm{edge} / \langle j \rangle$ are calculated. As the plasma moves downward toward the X-point between $t = 3680$~ms and $t = 3720$~ms, the magnetic axis descends from $Z_\mathrm{axis} \approx -0.02$~m to approximately $-0.09$~m. During this motion, the plasma volume decreases, leading to a rise in edge temperature, a decrease in resistivity, and a narrowing of the skin depth from $w_r \approx 3.43$~cm to $\approx 3.32$~cm. The narrower skin depth concentrates the compression-induced current into a thinner layer, increasing the local current density. The $\alpha$ and $j_\mathrm{edge} / \langle j \rangle$ increase, pushing the plasma toward the P-B stability boundary.

The modeled trajectory from the toy model (green line) starts in the stable region and moves toward the upper right as the plasma compresses, crossing the marginal-stability boundary near $t \sim 3709$~ms (red star), close to the observed ELM identified from the $D_\alpha$ signal shown in the right panels. The trajectory shows the plasma continuing to venture deeper into the unstable region, although the adiabatic assumption is no longer valid after an ELM due to rapid particle and energy transport.

Figure~\ref{fig:growthrate_3680} presents the time evolution of key parameters for the same $t = 3680$~ms cycle: the resistive skin depth $w_r$, the normalized edge current density, and the normalized growth rate $\gamma/(\omega_*/2)$. The resistive skin depth narrows from $w_r \approx 3.43$~cm at the cycle top to $\approx 3.32$~cm at the ELM, as the adiabatic compression raises the edge temperature and lowers the resistivity. The normalized edge current density rises as the induced current is concentrated into the narrowing layer. The normalized growth rate $\gamma/(\omega_*/2)$ from ELITE calculations begins near zero at the reference equilibrium, deep in the stable region, and rises over the downstroke, crossing the marginal-stability threshold into the unstable region within a few milliseconds of the ELM. Beyond the ELM the traces are faded, where the adiabatic model no longer correctly describes the post-ELM evolution.

Nonetheless, the trajectory demonstrates that the increase in edge current density from the compression and motion during the jog pushes the plasma across the stability boundary and further into the unstable region. The model captures the essential physics: during the downward motion, the narrowing skin depth amplifies the edge current perturbation contributing to destabilization of the ETB. After the ELM at $t \approx 3709$~ms, the plasma begins to recover and move upward. The modeled trajectory shows the operating point retreating back toward the stable region as the edge current relaxes and the pressure gradient decreases. This recovery sets the stage for the next jog cycle. 

This evolution is observed across the jog cycles in 174848 as well as in 174863, 174875, and 174862. Figure~\ref{fig:stability_timetrace} shows the analysis applied to six jog cycles of 174848. The jog drives up the normalized edge current $j_\mathrm{edge}/\langle j\rangle$ and growth rate $\gamma/(\omega_*/2)$ across the peeling boundary on each downstroke. The trajectories cross into and venture deeper in the unstable region followed by an ELM within a few milliseconds. Shot 174863 has downward jogs at different phases of the ELM cycle, i.e., jogs are applied to different pedestal and current profiles. Shot 174875 has upward jogs at different phases of the ELM cycle. From these jogs, the volume increases marginally compared to decreases observed in downward jogs. The upward motion drives current in the other direction, i.e., the same direction as the plasma current. The motion term dominates over the expansion term, thus driving a net positive current to trigger the ELM. Shot 174862 has jogs of varying displacement and thus volume change. Across all of these shots, ELMs are triggered according to the edge current evolution from the toy model. Additional analyses of 174863, 174875, and 174862 are in Appendices A, B, and C, respectively.

\section{Discussion}
\label{sec:discussion}
The ELM-jog phase analysis shows that ELMs in DIII-D are preferentially triggered during the downward motion of the LSN plasma toward the X-point. This is consistent with observations at JET and AUG \cite{de_la_luna_understanding_2016, lang_frequency_2004}, but contrasts with experiments at TCV, NSTX, and KSTAR where triggering occurred when the plasma moved away from the X-point \cite{degeling_magnetic_2003, gerhardt_first_2010, kim_elm_2012}. The DIII-D, JET, and AUG experiments were conducted in standard type-I ELMy H-mode regimes, whereas the TCV discharges exhibited characteristics of type-III ELMy H-modes with lower edge temperatures \cite{degeling_magnetic_2003}. The difference in triggering direction is related to the different coil configurations and vertical stability control systems across these devices, which affect the spatial structure of the external flux perturbation $\delta\psi_\mathrm{ext}$ and plasma shape during a jog. In terms of the induced-current decomposition of Eq.~(\ref{eq:edge_current}), the toy model contains both a compression contribution ($-B_\theta(r_0) R_0\,\delta w_r$) and a motional contribution ($\delta\mathbf{r}\cdot\nabla\psi_{ext}$), and the triggering phase is set by which one dominates for a given configuration and jog direction. In DIII-D shot 174848 with LSN plasma and the downward jogs, motion toward the X-point compresses the cross-section which dominates edge current evolution. Triggering by upward motion, as observed in DIII-D shot 174875, corresponds instead to the motional $\delta\mathbf{r}\cdot\nabla\psi_{ext}$ term dominating, which can drive the edge current positive on the stroke that carries the plasma away from the X-point.

The higher-than-natural ELM frequency produced by jogging resulted in significantly smaller individual ELM energy losses, consistent with the inverse relationship between ELM frequency and ELM size: the fractional stored-energy loss dropped from above 10\% (up to $\sim$25\%) for the natural ELMs to less than $\sim$8\% for the jog-triggered ELMs. This directly reduced the peak divertor heat flux by approximately a factor of 2. The simultaneous reduction in $Z_\mathrm{eff}$ indicates enhanced impurity pump-out from the more frequent ELMs, which is particularly relevant for future burning plasma devices with tungsten plasma-facing components where controlling impurity accumulation is critical for avoiding radiative collapse \cite{Philipps_2011}.

The ELM triggering mechanism of jogging in the DIII-D experiments can be attributed to the induction of an edge toroidal current, consistent with the simulations and observations\cite{artola_non-linear_2018, de_la_luna_understanding_2016, de_la_luna_magnetic_2009}. When the plasma moves rapidly downward toward the X-point, the plasma volume is compressed. This compression induces a net positive toroidal current that redistributes the edge electromagnetic profile, pushing the plasma state across the peeling side of the P-B stability boundary into the unstable region, thereby triggering an ELM \cite{Connor_1998, snyder_edge_2002}. The toy model reproduces the timing of the stability boundary crossing to within a few milliseconds of the observed ELM onset, supporting this picture.

During these DIII-D experiments, 3 to 4 compound ELMs were produced per vertical jog. This compound triggering behavior is similar to observations at KSTAR, which produced 2 to 3 ELMs per jog while oscillating between LSN and upper single null (USN), and transiently at ASDEX Upgrade, which produced 1 to 2 ELMs per jog \cite{kim_elm_2012, lang_frequency_2004}. The compound ELM phenomenon can be understood in the context of the toy model and the P-B stability diagram. The timescale of the vertical jog ($\sim$50~ms per stroke) is significantly longer than the timescale of an individual ELM crash ($\lesssim$1~ms). Consequently, the jog-driven compression and associated edge current induction persist well after the first ELM. The first ELM partially relaxes the pedestal pressure gradient and edge current, moving the operating point back into the stable region. However, because the plasma is still being compressed during the ongoing downward motion, the edge current continues to be driven by the mechanism described in Section~\ref{sec:result}, pushing the operating point back across the stability boundary before the pedestal fully recovers. The number of compound ELMs per jog is thus determined by the competition between the rate of edge current induction and the rate of pedestal relaxation after each ELM crash. The low collisionality of the DIII-D type-I ELMy H-mode regime may also contribute: at lower collisionality the induced edge current decays more slowly (lower resistivity, $\eta \propto T_e^{-3/2}$) and, because the pedestal bootstrap current is less collisionally suppressed, more edge current is produced at a given pressure gradient so the pedestal operates closer to the low-$n$ peeling stability limit~\cite{snyder_elms_2004, snyder_lowcoll_2007}, both of which favor multiple re-crossings of the stability boundary per jog cycle.

Practically, vertical jogs offer a distinct advantage over other ELM control techniques in that they do not require additional dedicated hardware. All tokamaks already have poloidal field coils for vertical stability control, making jogging a readily available fallback technique. This is particularly relevant for ITER, where vertical oscillations have been considered as a backup ELM control method for low-current ($\sim$7.5~MA) operation scenarios \cite{gribov_plasma_2015, lang_elm_2013}. At higher plasma currents ($\sim$15~MA, $Q = 10$ operation), the achievable vertical displacement is limited to 2--3~cm by the vertical stability control system, which may reduce the effectiveness of jogging. Nevertheless, for scenarios where other ELM control methods such as RMPs or pellet injection are unavailable or insufficient, vertical jogs provide a viable alternative.

The agreement between the modeled stability crossing and the observed ELM timing provides evidence that the mechanism of ELM triggering by vertical jogs is indeed the induction of edge toroidal current through plasma compression in an inhomogeneous magnetic field.

\section{Conclusion}
\label{sec:conclusion}
ELM pacing via vertical plasma oscillations, or jogging, has been successfully demonstrated on the DIII-D tokamak. Rapid vertical movements of the plasma at 10~Hz each triggered 3--4 compound ELMs, raising the effective ELM frequency from $\sim$5~Hz to above 20~Hz. ELMs were found to occur preferentially during the downward phase of the jog in the LSN configuration. The higher ELM frequency reduced individual ELM energy losses to below $\sim$8\% and the peak divertor heat flux by a factor of around 2. In addition, the reduction of $Z_\mathrm{eff}$ shows the higher frequency ELMs were effectively pumping out impurities. P-B stability analysis has been conducted using a toy model of edge current evolution and ELITE. The results indicate that compression-driven edge current induction is the primary mechanism of ELM triggering by jogging.

\section*{Acknowledgments}
This material is based upon work supported by the U.S. Department of Energy, Office of Science, Office of Fusion Energy Sciences, using the DIII-D National Fusion Facility, a DOE Office of Science user facility, under Award(s) DE-FC02-04ER54698, DE-AC02-09CH11466, DE-SC0022270.

\section*{Disclaimer}
This report is prepared as an account of work sponsored by an agency of the United States Government. Neither the United States Government nor any agency thereof, nor any of their employees, makes any warranty, express or implied, or assumes any legal liability or responsibility for the accuracy, completeness, or usefulness of any information, apparatus, product, or process disclosed, or represents that its use would not infringe privately owned rights. Reference herein to any specific commercial product, process, or service by trade name, trademark, manufacturer, or otherwise, does not necessarily constitute or imply its endorsement, recommendation, or favoring by the United States Government or any agency thereof. The views and opinions of authors expressed herein do not necessarily state or reflect those of the United States Government or any agency thereof.

\appendix
\section{Shot 174863 with jogs at different phases of ELM cycle}
\label{app:174863}
\begin{figure}[htbp]
\centering
\includegraphics[width=\textwidth, keepaspectratio]{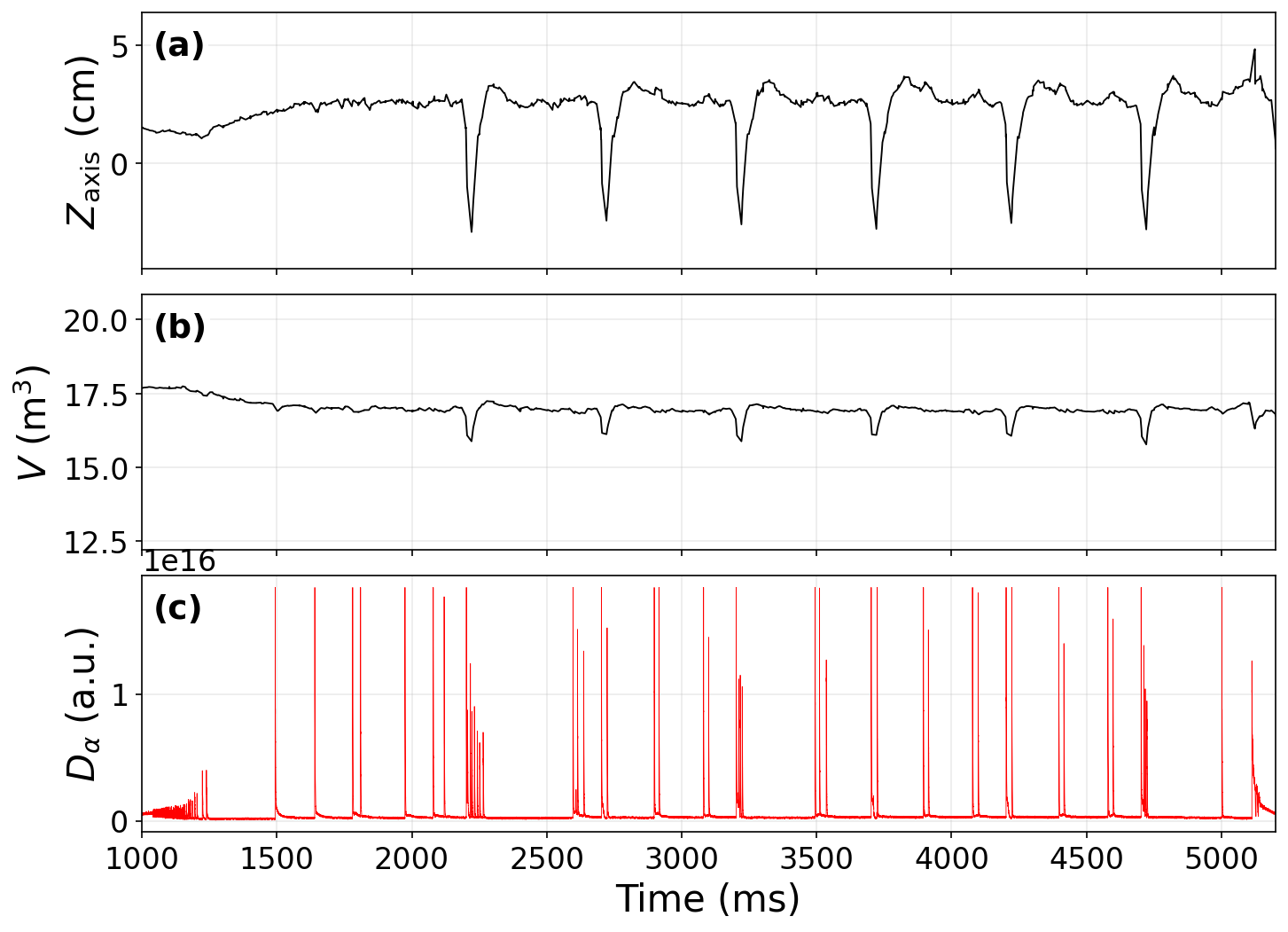}
    \caption{Overview of DIII-D shot 174863 with key parameters: (a) magnetic-axis vertical position $Z_\mathrm{axis}$, (b) plasma volume $V$ , and (c) $D_\alpha$.}
    \label{fig:overview_174863}
\end{figure}

\begin{figure}[htbp]
\centering
\includegraphics[width=\textwidth, keepaspectratio]{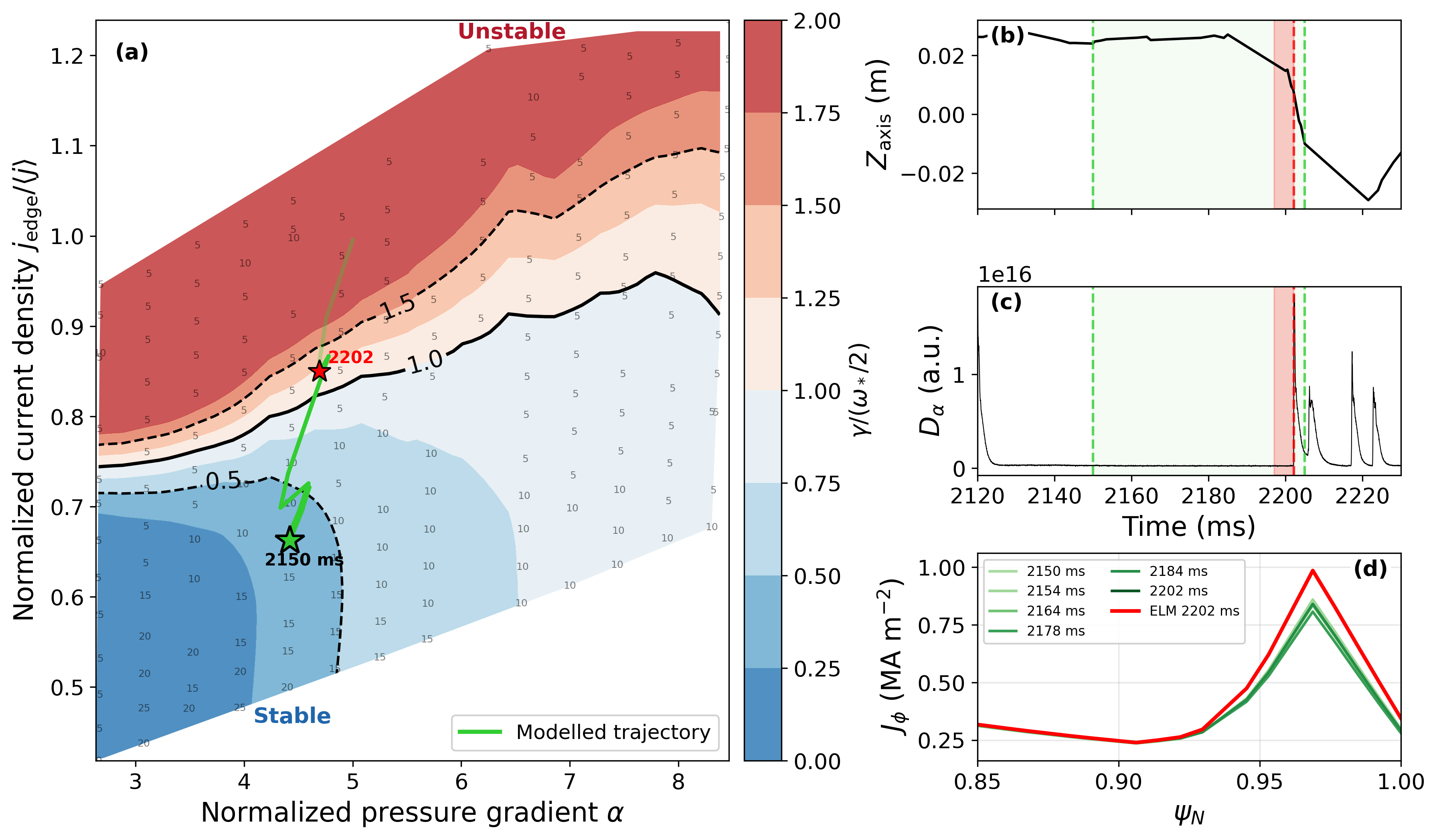}
    \caption{P-B stability diagram for DIII-D shot 174863 at $t = 2150$~ms, analogous to Figure~\ref{fig:ELITE}. (a) The $\gamma/(\omega_*/2)$ contour map is shown in the ($\alpha$, $j_\mathrm{edge}/\langle j\rangle$) plane. The green star marks the reference equilibrium ($t = 2150$~ms), the green curve the modeled edge operating-point trajectory (faded after the ELM), and the red star the ELM onset. (b) Magnetic-axis position $Z_\mathrm{axis}$, (c) the $D_\alpha$ signal, and (d) the modeled edge toroidal current-density profiles $J_\phi(\psi_N)$ at successive points along the trajectory are shown over the same window.}
    \label{fig:ELITE_174863}
\end{figure}

\begin{figure}[htbp]
\centering
\includegraphics[width=0.58\linewidth, height=0.72\textheight, keepaspectratio]{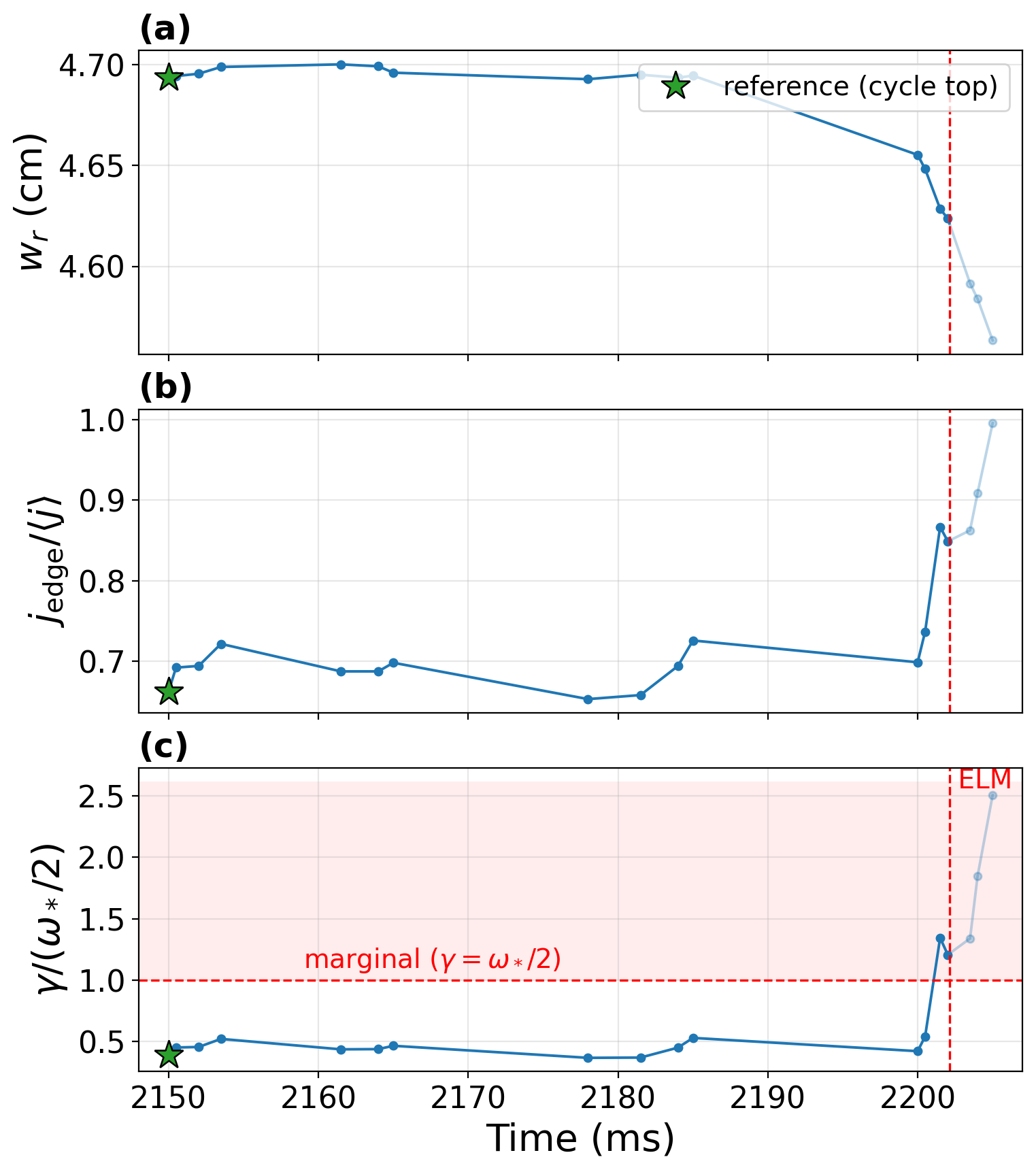}
    \caption{Time evolution over the modeled $t = 2150$~ms downstroke of shot 174863, in the format of Figure~\ref{fig:growthrate_3680}: (a) the resistive skin depth $w_r$, (b) the normalized edge current density $j_\mathrm{edge}/\langle j\rangle$, and (c) the normalized growth rate $\gamma/(\omega_*/2)$. }
    \label{fig:growthrate_174863}
\end{figure}

The same toy-model and ELITE workflow applied to 174848 has been repeated for 174863. Figure~\ref{fig:overview_174863} gives a full-shot overview of shot 174863. Figure~\ref{fig:ELITE_174863} shows the P-B stability diagram for shot 174863 at $t = 2150$~ms. As in 174848, the compression-driven edge current evolution carries the operating point from the reference equilibrium across the peeling boundary at the ELM, consistent with the mechanism described in Section~\ref{sec:result}. The corresponding time traces of the skin depth $w_r$, the normalized edge current density $j_\mathrm{edge}/\langle j\rangle$, and the growth rate $\gamma/(\omega_*/2)$ over this downstroke are shown in Figure~\ref{fig:growthrate_174863}.

\section{Shot 174875 with upward jogs}
\label{app:174875}
\begin{figure}[htbp]
\centering
\includegraphics[width=\textwidth, keepaspectratio]{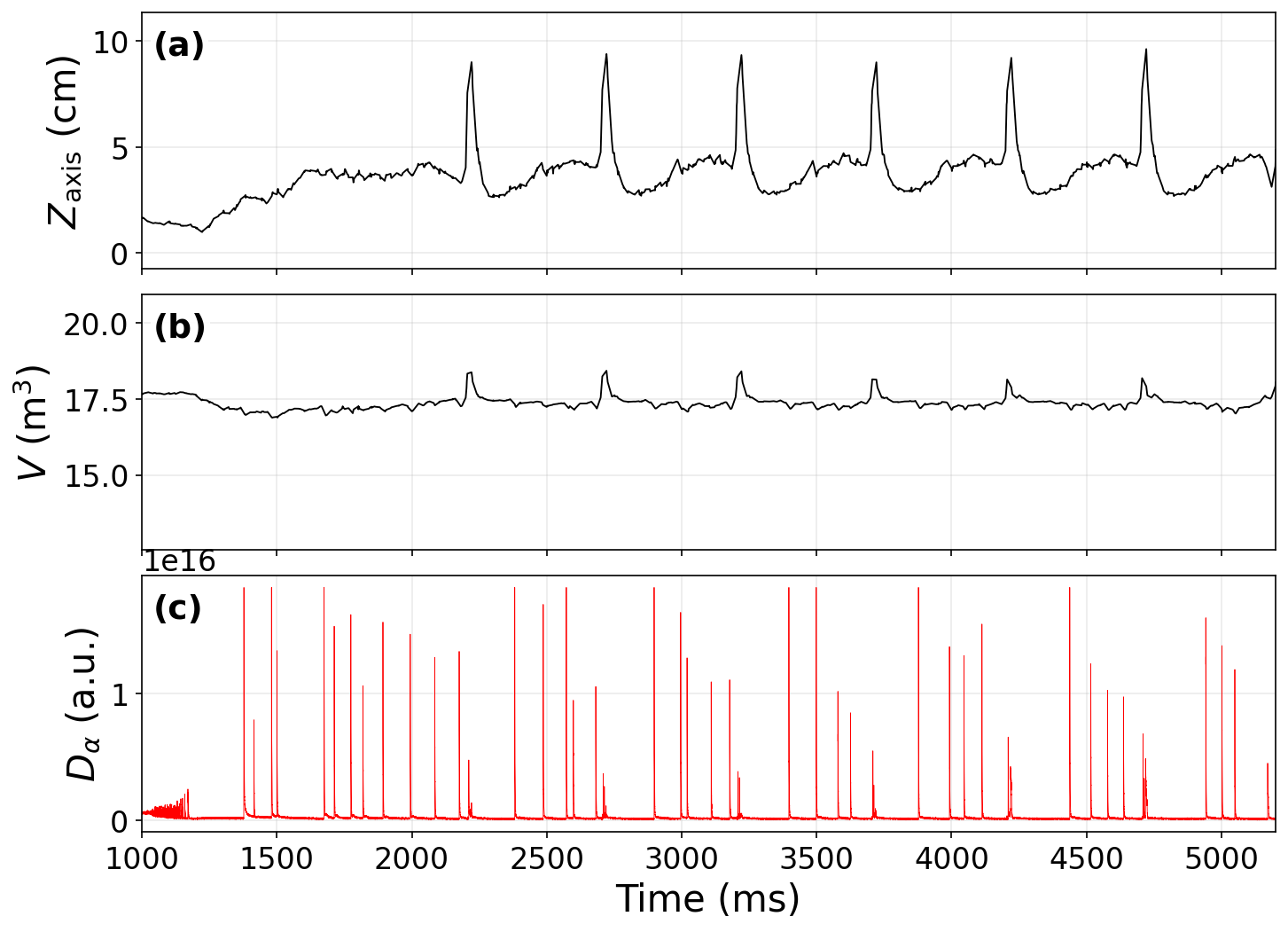}
    \caption{Overview of DIII-D shot 174875 with key parameters: (a) magnetic-axis vertical position $Z_\mathrm{axis}$, (b) plasma volume $V$, and (c) the $D_\alpha$. This is the upstroke-triggered discharge.}
    \label{fig:overview_174875}
\end{figure}

\begin{figure}[htbp]
\centering
\includegraphics[width=\textwidth, keepaspectratio]{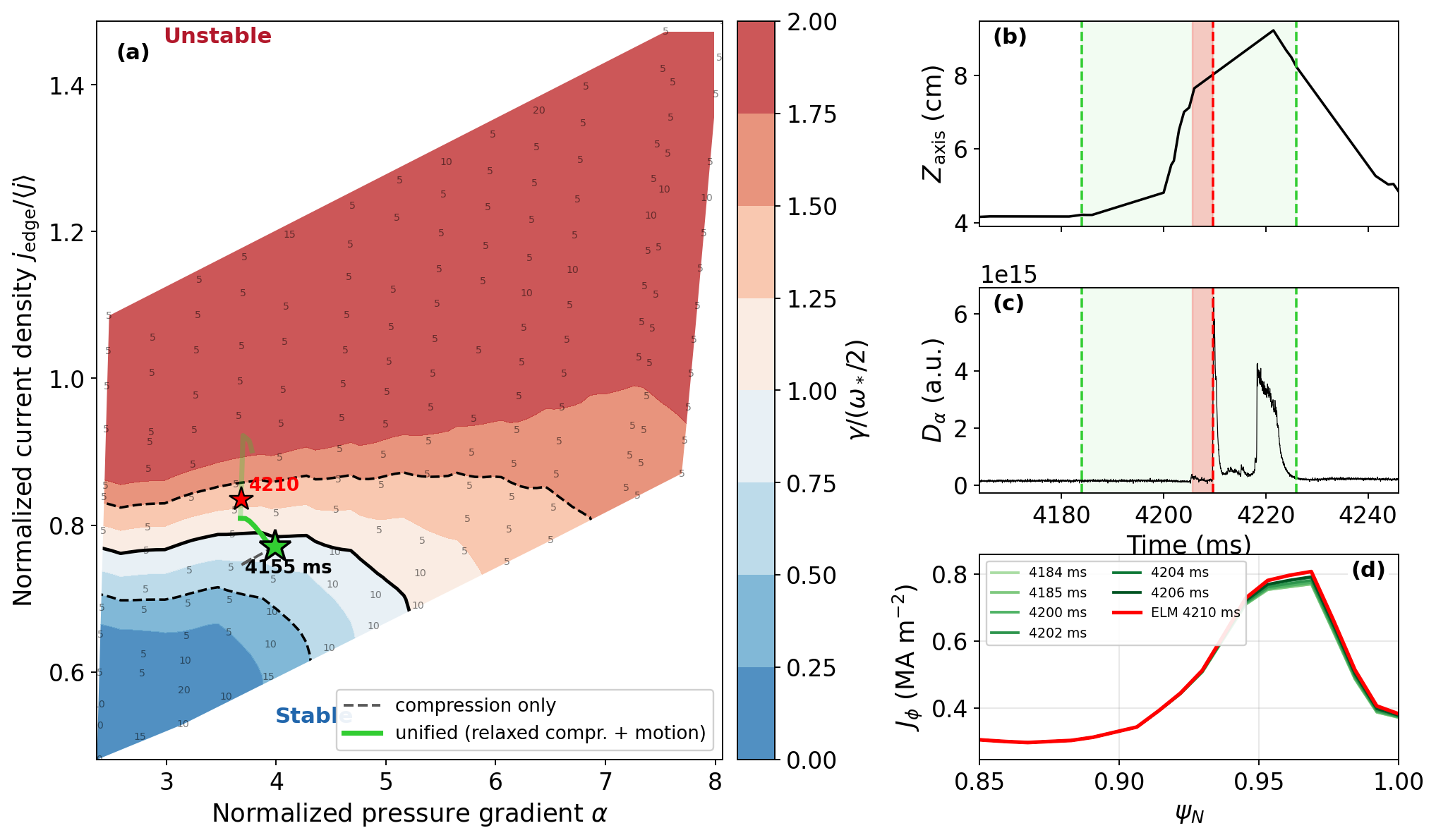}
    \caption{P-B stability evolution for DIII-D shot 174875 (upstroke-triggered), anchored at $t = 4155$~ms. (a) The $\gamma/(\omega_*/2)$ contour map is shown in the ($\alpha$, $j_\mathrm{edge}/\langle j \rangle$) plane: the green star marks the reference equilibrium and the solid green curve the modeled operating-point trajectory during the upward jog, driven by the motiona term $\delta\mathbf{r}\cdot\nabla\psi_{ext}$; the red star marks the ELM onset. (b) $Z_\mathrm{axis}$, (c) $D_\alpha$ timetraces, with the simulation window (green) and ELM onset (red) marked, (d) the edge toroidal current-density profile $J_\phi(\psi_N)$ evolving from the reference are shown.}
    \label{fig:ELITE_174875}
\end{figure}

\begin{figure}[htbp]
\centering
\includegraphics[width=0.58\linewidth, height=0.72\textheight, keepaspectratio]{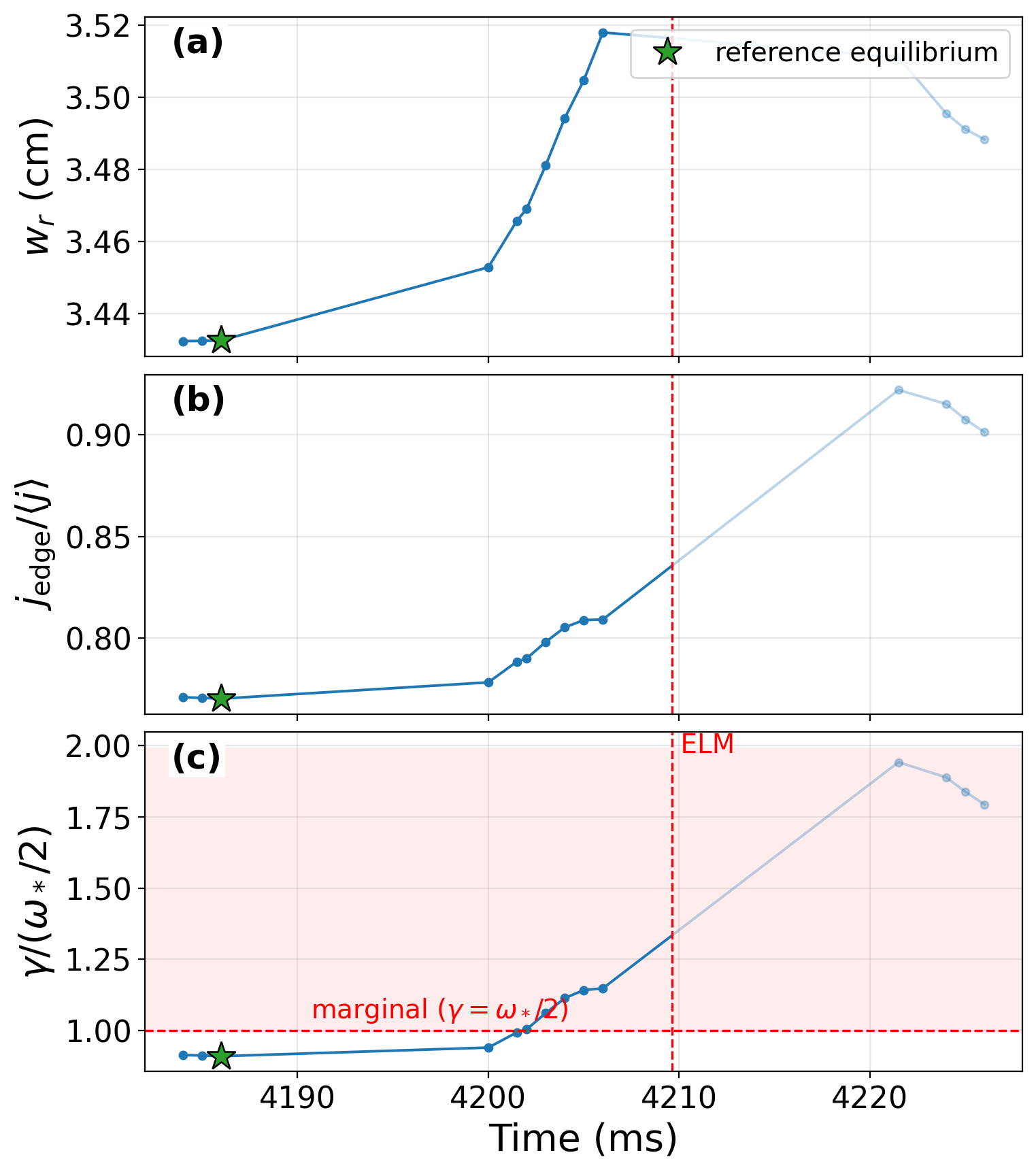}
    \caption{Time evolution over the modeled $t = 4155$~ms upstroke of shot 174875 with key parameters: (a) the resistive skin depth $w_r$, (b) the normalized edge current density $j_\mathrm{edge}/\langle j\rangle$, and (c) the normalized growth rate $\gamma/(\omega_*/2)$. The green star marks the reference equilibrium and the red dashed line the observed ELM onset. The traces are faded after the ELM.}
    \label{fig:growthrate_174875}
\end{figure}

Shot 174875 is a case, in which ELMs are also observed during the \emph{upward} phase of the jog. Figure~\ref{fig:overview_174875} gives an overview. Figure~\ref{fig:ELITE_174875} shows the stability evolution anchored at $t = 4155$~ms. In this discharge the plasma expands on the upstroke, so the compression ($\delta w_r$) term is stabilizing; the destabilizing drive instead comes from the motional $\delta\mathbf{r}\cdot\nabla\psi_{ext}$ contribution, which becomes large as the magnetic axis rises rapidly through the sheared external field. Note the reference equilibrium of this discharge already sits closer to the peeling boundary than that of reference time points from other discharges. Nonetheless, the same edge-current model---now with the motion induced current rather than the compression providing the drive---accounts for upstroke triggering. Figure~\ref{fig:growthrate_174875} shows the corresponding time traces of the skin depth $w_r$, the edge current density $j_\mathrm{edge}/\langle j\rangle$, and the growth rate $\gamma/(\omega_*/2)$ over this upstroke: the skin depth \emph{widens} as the plasma expands, yet the motional term drives the edge current density and the growth rate up across the marginal-stability boundary in coincidence with the observed ELM.

\section{Shot 174862 with jogs of different sizes}
\label{app:174862}
\begin{figure}[htbp]
\centering
\includegraphics[width=\textwidth, keepaspectratio]{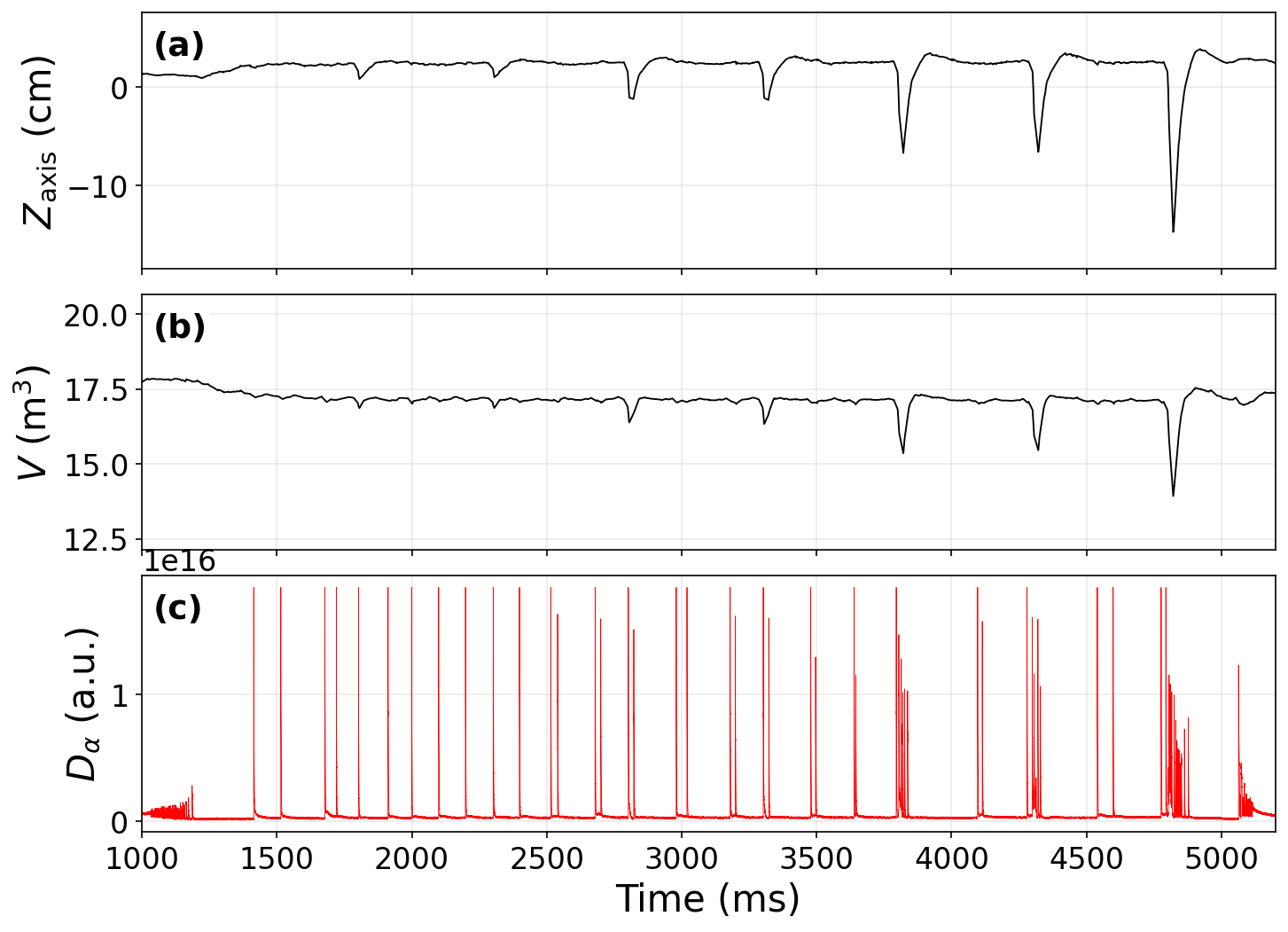}
    \caption{Overview of DIII-D shot 174862 with key parameters: (a) magnetic-axis vertical position $Z_\mathrm{axis}$, (b) plasma volume $V$, and (c) the $D_\alpha$.}
    \label{fig:overview_174862}
\end{figure}

\begin{figure}[htbp]
\centering
\includegraphics[width=\textwidth, keepaspectratio]{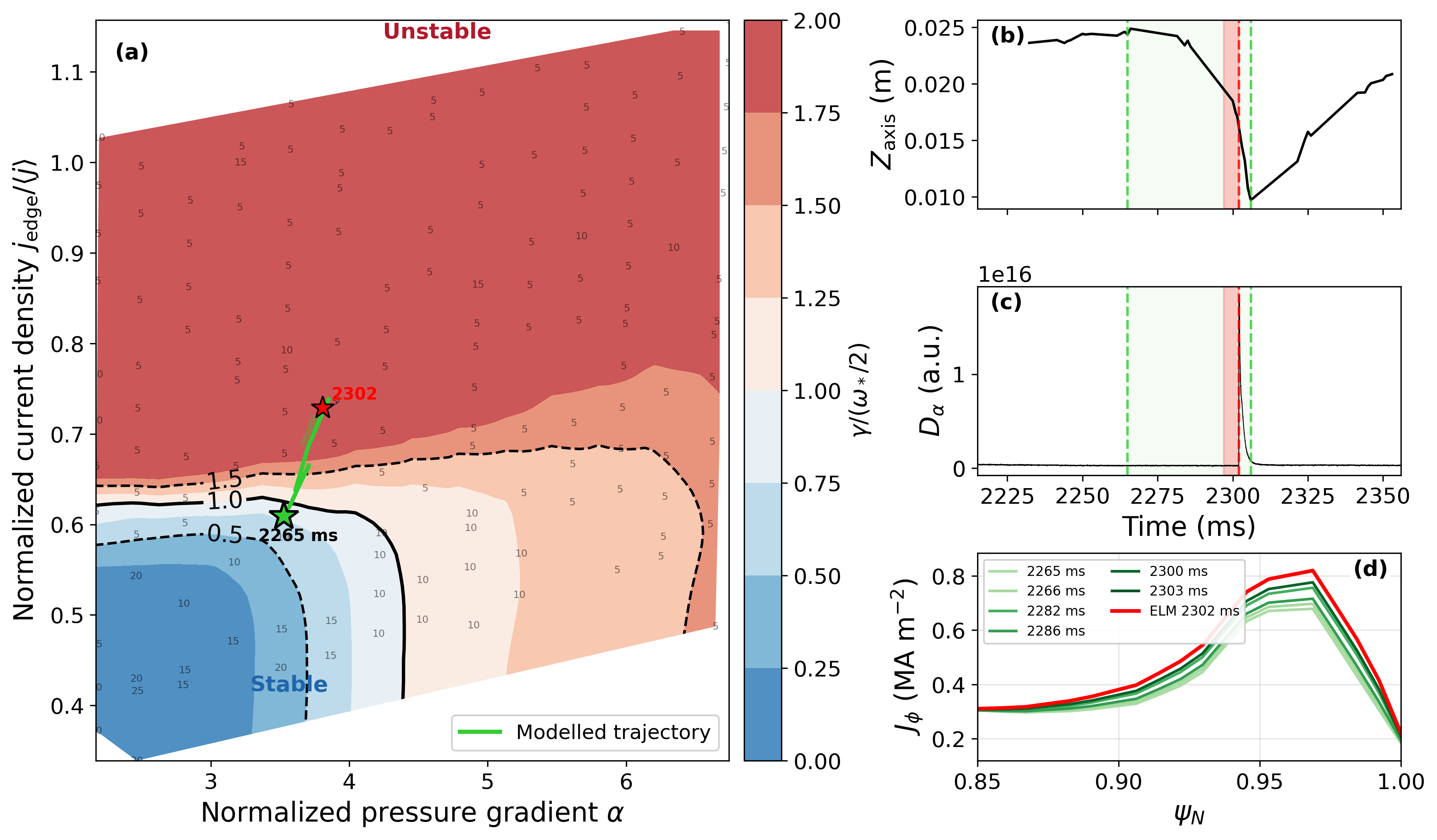}
    \caption{P-B stability diagram for DIII-D shot 174862 at $t = 2265$~ms with key parameters: (a) The $\gamma/(\omega_*/2)$ contour map is shown in the ($\alpha$, $j_\mathrm{edge}/\langle j\rangle$) plane. The green star marks the reference equilibrium ($t = 2265$~ms), the green curve the modeled edge operating-point trajectory (faded after the ELM), and the red star the ELM onset. (b) Magnetic-axis position $Z_\mathrm{axis}$, (c) the $D_\alpha$ signal, and (d) the modeled edge $J_\phi(\psi_N)$ profiles along the trajectory are shown over the same window.}
    \label{fig:ELITE_174862}
\end{figure}

\begin{figure}[htbp]
\centering
\includegraphics[width=0.58\linewidth, height=0.72\textheight, keepaspectratio]{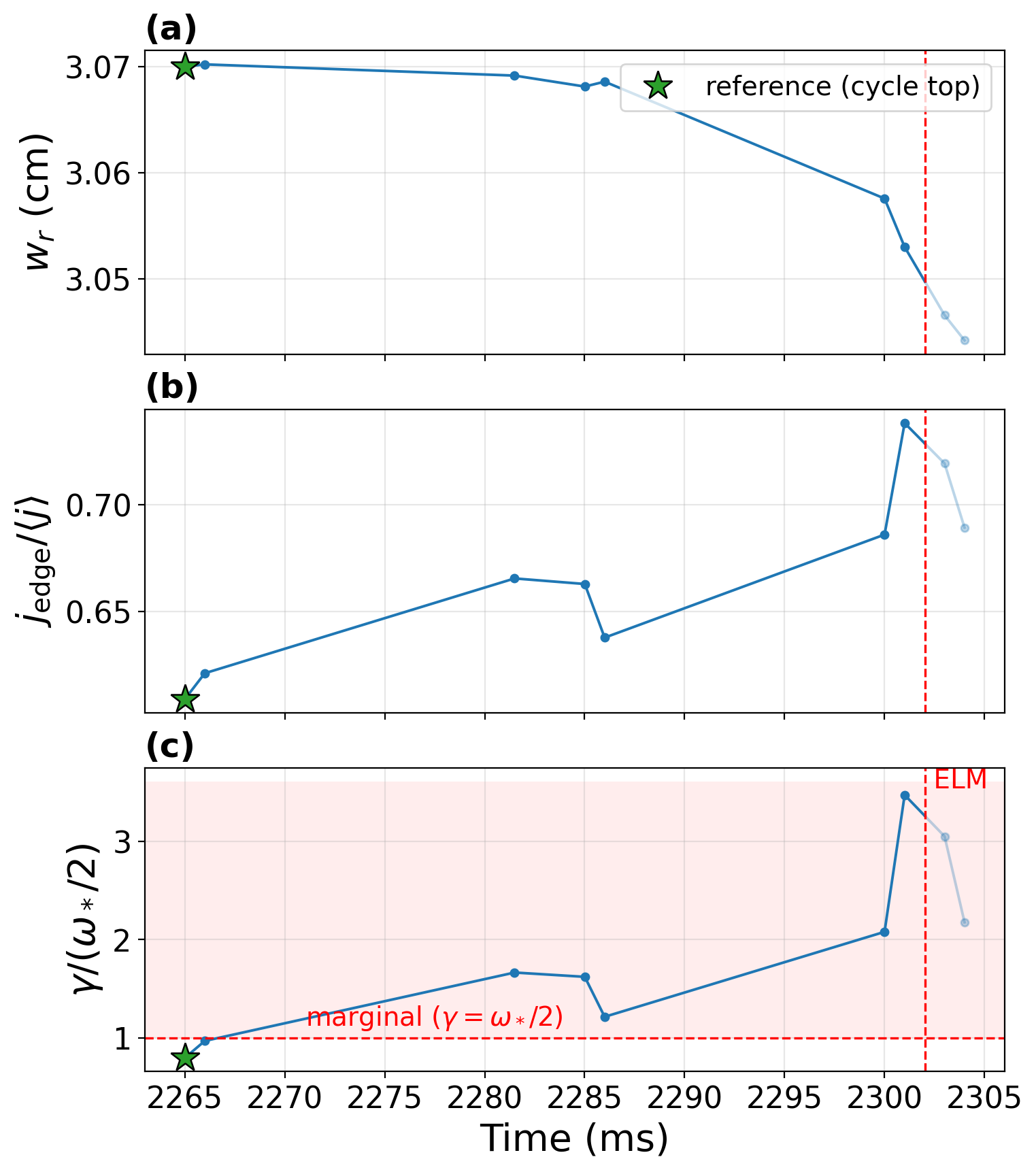}
    \caption{Time evolution over the modeled $t = 2265$~ms jog of shot 174862, in the format of Figure~\ref{fig:growthrate_3680}: (a) the resistive skin depth $w_r$, (b) the normalized edge current density $j_\mathrm{edge}/\langle j\rangle$, and (c) the normalized growth rate $\gamma/(\omega_*/2)$. The green star marks the reference equilibrium and the red dashed line the observed ELM onset. The traces are faded after the ELM.}
    \label{fig:growthrate_174862}
\end{figure}

The same toy-model and ELITE workflow used for the other discharges has been applied to shot 174862, anchored at the reference equilibrium at $t = 2265$~ms. Figure~\ref{fig:overview_174862} gives an overview. Figure~\ref{fig:ELITE_174862} shows the P-B stability diagram. Figure~\ref{fig:growthrate_174862} shows the corresponding time traces of the skin depth $w_r$, the normalized edge current density $j_\mathrm{edge}/\langle j\rangle$, and the growth rate $\gamma/(\omega_*/2)$. ELMs were triggered regardless of the jog size.

\begin{figure}[htbp]
\centering
\includegraphics[width=\textwidth, keepaspectratio]{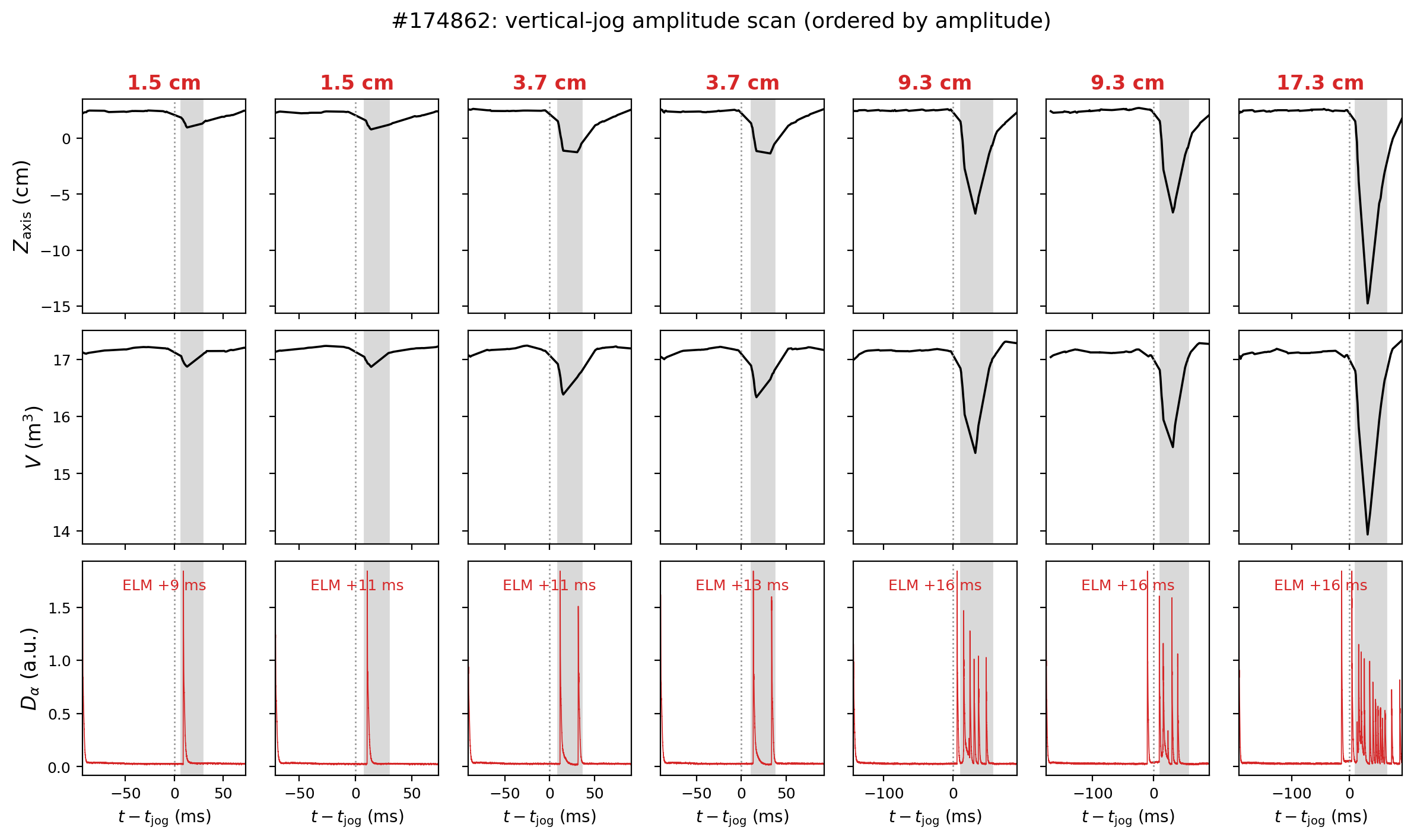}
    \caption{Vertical-jog amplitude scan in DIII-D shot 174862. Each column is one isolated downward jog, ordered left-to-right by increasing amplitude. In each column the time axis is referenced to the jog onset. Top row: magnetic-axis vertical position; middle row: plasma volume $V$; bottom row: divertor $D_\alpha$, with the shaded band marking the compression window. Every sampled jog triggers at least one ELM locked to the compression. The number of ELMs per jog increases with amplitude---a single ELM at the smallest amplitudes, rising to bursts of several compound ELMs at the largest---as does the depth of the volume compression.}
    \label{fig:amplitude_scan}
\end{figure}

 Figure~\ref{fig:amplitude_scan} shows the magnetic-axis excursion, the plasma volume, and the divertor $D_\alpha$ signal in a short window about each jog, ordered by amplitude. Every jog in the scan triggers at least one ELM locked to the compression window. The plasma volume is compressed by an amount that grows with amplitude, and the number of ELMs triggered per jog is observed to rise correspondingly. The amplitude threshold for triggering is therefore not resolved within the sampled range, but the induced-current picture requires a minimum displacement to drive the edge operating point across the peeling boundary. Such a threshold is therefore expected to lie below the smallest amplitude sampled here.

\bibliography{bib_elm_pacing}

\end{document}